\documentclass[prb,onecolumn,preprint]{revtex4-1}
\usepackage[latin1]{inputenc}
\usepackage{hyperref}
\usepackage{amsmath}
\usepackage{graphicx}
\usepackage{graphics}
\usepackage{amsfonts}
\usepackage{amssymb}
\usepackage{natbib}
\usepackage{subfigure}
\newcommand\sro{Sr$_2$RuO$_4$}
\newcommand\tc{T$_{\rm c}$}
\newcommand\upt{UPt$_3$}
\newcommand\he{$^3$He}
\begin{document} 
\title{Chiral Superconductors}
\date{\today}
\author{Catherine Kallin}
\affiliation{McMaster University, Hamilton, ON, Canada}
\author{John Berlinsky}
\affiliation{Kavli Institute for Theoretical Physics, Santa Barbara, CA}
\begin{abstract}

Chiral superconductivity is a striking quantum phenomenon in which an unconventional superconductor spontaneously develops an angular momentum and lowers its free energy by eliminating nodes in the gap.  It is a topologically non-trivial state and, as such, exhibits distinctive topological modes at surfaces and defects. In this paper we discuss the current theory and experimental results on chiral superconductors, focusing on two of the best-studied systems, \sro, which is thought to be a chiral triplet p-wave superconductor, and \upt, which has two low-temperature superconducting phases (in zero magnetic field), the lower of which is believed to be chiral triplet f-wave.  Other systems that may exhibit chiral superconductivity are also discussed.  Key signatures of chiral superconductivity are surface currents and chiral Majorana modes, Majorana states in vortex cores, and the possibility of half-flux quantum vortices in the case of triplet pairing.  Experimental evidence for chiral superconductivity from $\mu$SR, NMR, strain, polar Kerr effect and Josephson tunneling experiments are discussed.
\end{abstract}
\maketitle
\section{Introduction}
\subsection{Overview}
Superconductivity has passed through many phases or eras in its history.  Starting from its discovery in 1911\cite{ko}, there was the long theoretical desert, lasting 46 years, when there was no microscopic theory, which was followed by the golden age of BCS theory\cite{bcs} when it seemed that everything was worked out\cite{parks}.  This led to predictions of ``anisotropic superfluidity"\cite{am,bw} which were later realized in the various superfluid phases of liquid $^3$He at milliKelvin temperatures.\cite{orl, wheat, leggett,leggettbook}  Still everything seemed understood until the bombshell discovery of high \tc\ in 1985\cite{bm} where the new ingredient of strongly correlated electrons opened challenges for theorists and experimentalists which persist to today.  More recently the significance of the topological nature of various insulating and superconducting states has led to new insights and to the study of  particles, such as Majorana fermions, which were new to condensed matter physics\cite{beenakker,ivanov}.  The potential applications of such particles in the context of quantum computing\cite{npj} have raised the stakes for understanding topological superconductors and their excitations which we will address in this short review on the topic of ``chiral superconductors."

A chiral superconductor is one in which the phase of the complex superconducting gap function, $\Delta(\vec k)$, winds in a clockwise or counter-clockwise sense as $\vec k$ moves about some axis on the Fermi surface of the underlying metal.  The simplest example is a $k_x\pm i k_y$ gap function which precesses by $\pm 2\pi$ as $\vec k$ follows a closed path enclosing the $k_z$ axis.  A chiral gap function breaks time reversal symmetry and is degenerate with its time-reversed partner.  Chiral superconductivity is a type of topological state, and it carries with it certain signatures of its non-trivial topology.\cite{chiraltop}  For example, there are localized states at the edge which disperse across the gap for $\vec k$ parallel to the edge and perpendicular to the chiral axis, passing through zero energy for $\vec k = 0$.\cite{read}  These are called Majorana states, and the zero energy state at $\vec k =0$ has the Majorana property of being its own antiparticle.  Similar physics applies at vortex cores in chiral superconductors which for chiral p-wave exhibit a single Majorana zero mode for the case of spinless fermions.\cite{read}  These Majorana modes can be stable to local perturbations, protected by the bulk gap because of its topological nature.

Another striking property is the expected existence of surface currents in chiral superconductors, which are carried both by the edge modes and by the bulk states near the surface which are orthogonal to the edge modes.\cite{stone}  In the simplest theory of chiral p-wave, these edge currents are substantial, although they would be screened by counter-rotating diamagnetic currents.  Since the length scales for the surface currents and the screening currents are different, one would expect to see easily measurable spatially varying magnetic fields near the surface.  These edge currents, even without screening, are not topological or quantized, but the edge states and topology of the chiral superconducting state can give rise to a quantized thermal Hall conductance.\cite{senthil, read,newsigrist}

For triplet chiral superconductors, such as chiral p- or f-wave,  additional novel properties arise from the spin degrees of freedom.  Triplet superconductors are special. The vast majority of known superconductors are spin-singlet superconductors. This includes all conventional s-wave superconductors and the high \tc\ cuprates and pnictides. There are a relatively small number of candidates for spin-triplet superconductivity, several with substantial evidence pointing toward triplet pairing, but none that are as firmly established as the triplet nature of superfluid \he.\cite{leggett, leggettbook,wheat} Amongst the strongest candidates for triplet pairing are \upt\ and \sro, which are also the leading candidates for chiral superconductivity. Other possible triplet superconductors include ferromagnetic heavy fermion superconductors such as UGe$_2$\ and UCoGe, and noncentrosymmetric (lacking inversion symmetry) superconductors such as Li$_2$Pt$_3$B. For other examples of possible triplet superconductors, see Ref.~\onlinecite{maenotriplet}. 
Most of these materials have strong spin-orbit coupling and are more accurately described as odd-parity, pseudospin-triplet superconductors. However, here we will often drop the pseudo or pseudospin, and simply call them triplet superconductors. 

One of the exotic phenomena that can occur in triplet superconductors is half-quantum vortices that carry half the usual superconducting flux quantum.\cite{leggett,leggettbook,read} For chiral p-wave, each half-quantum vortex has a single Majorana zero mode in the vortex core, similar to the spinless chiral p-wave case, and the vortex-Majorana mode composites obey non-Abelian braiding statistics.\cite{ivanov} This means that one can move between distinct, topologically stable states by moving such vortices around each other\cite{ivanov} and makes the possible realization of chiral p-wave superconductivity in real materials of interest to the quantum computing community.\cite{npj}  Half-quantum vortices may have been observed in \sro.\cite{budakian}

\subsection{Why would a superconductor be chiral?}
\label{why}
The simplest superconductors are s-wave with a single Fermi surface, and the attractive interaction that supports pairing is the electron-phonon interaction.  For s-wave, the gap function has the full symmetry of the crystal.  When the interaction is repulsive at short distance but overall attractive, Cooper pairing may happen in some higher angular momentum channel in order to avoid the short-range repulsion.  The classic example is superfluid \he\ where the interaction is the He-He potential and the Cooper pairs have angular momentum 1 (p-wave).\cite{leggett,wheat}  The high-T$_c$ cuprates are d-wave,\cite{dwave} presumably because of the strong on-site Coulomb repulsion and longer range attractive exchange interactions.

A distinctive property of most non-s-wave pairing is the presence of nodes in the gap.  For example, the $d_{x^2-y^2}$ gap function in the cuprates has nodes along $k_x=\pm k_y$.  A hypothetical $p_x$ superconductor would have nodes along $p_y$ and a $p_x+p_y$ superconductor would have nodes along $k_x=-k_y$.  Since the condensation energy of a superconductor depends on the gap squared, $|\Delta(\vec k)|^2$, averaged over the Fermi surface, nodes are generically bad for superconductivity.  Nevertheless they happen when there is a good fit between a particular type of gap with nodes and the underlying microscopics.  Sometimes it happens, however, that two equivalent gap functions are degenerate, for example, $p_x$ and $p_y$ in a tetragonal crystalline system or in an isotropic liquid like $^3He$.  In that case, one can form linear combinations such as $p_x\pm ip_y$ whose absolute value squared is isotropic. For a cylindrical Fermi surface, this state is nodeless which optimizes the condensation energy for p-wave.  A similar situation occurs in hexagonal systems where $d_{x^2-y^2}$ and $d_{xy}$ are also degenerate. Of course, the system must choose either $+$ or $-$ in the $\pm i$, thus determining the chirality of the state.  This choice is a spontaneous symmetry breaking and the result is a chiral superconductor of the type discussed above. In principle, domains of the two chiralities may coexist at the cost of some interfacial or domain wall energy.\cite{domain1,domain2} Such domain walls, once created, can be pinned and stabilized by disorder. In a perfect sample, one expects only a single domain.

Another distinctive feature of materials which are candidate chiral superconductors is the presence of multiple Fermi surfaces, requiring multiple gap functions to describe the superconducting state.  The candidate chiral superconductors that will be discussed in this paper,  such as \sro\ and \upt, have multiple bands crossing the Fermi energy. Different bands exhibit different physical properties.  For example, in the quasi-2D material, \sro, two of the bands are quasi-1D along $\hat x$ or $\hat y$, while the third is rather isotropic in the $xy$\ plane.  An important question, for this system, is whether it is the 1D bands or the 2D bands that primarily stabilize superconductivity or whether all three bands participate roughly equally.  It is believed that multi-band effects are important for understanding the surface currents and in making chirality easier to detect, as will be discussed below. The presence of multiple bands and gaps enriches the possible structure of multi-band chiral superconductors.

\subsection{Plan for this paper}
In Section \ref{sec2} of this paper we discuss microscopic models
which include multiple bands, spin-orbit coupling and pairing into singlet and triplet superconducting states, as well as the formalism and nomenclature used for triplet superconductors.
In Section \ref{sec3}, we look at 1D and 2D lattice and continuum models with edges and vortices at the level of Bogolibov-deGennes (BdG) theory.  The focus here is on the zero-energy modes localized at chain ends and in vortex cores and the resulting bands of Majorana edge states that occur in chiral 2D systems.  We discuss why these low-energy states are Majorana modes and how they connect to observables in chiral superconductors, such as surface currents. Half-quantum vortices, which may support Majorana modes in a chiral superconductor, are also discussed in this Section. 

The polar Kerr effect, another consequence of chiral symmetry breaking which is proportional to the anomolous quantum Hall conductivity, is discussed in Section \ref{secKerr}.  
In Section \ref{secSRO} we describe the most important physical properties of \sro\ and how well they compare to our expectations for a chiral p-wave superconductor.  In Section \ref{secUPt} we discuss \upt\ which is a possible chiral f-wave superconductor, while in Section \ref{sec7} other possible chiral superconductors are discussed.  In Section \ref{sec8} we provide some final thoughts on our current state of understanding and the outlook for chiral superconductivity. 

\section{Microscopic models for chiral superconductors}
\label{sec2}
In many materials superconductivity arises from phonons, but in the case of higher angular momentum pairing, such as occurs in chiral superconductors, it is reasonable to consider purely electronic mechanisms.  These can arise from Coulomb interactions on a single, multi-orbital ion, although longer range Coulomb interactions may also play a role.  Starting from these microscopic interactions, weak-coupling and functional renormalization group methods have been used to try to understand the nature and origin of the pairing mechanisms. These calculations are discussed in the section on \sro.   Using such calculations as a guide, one can go on to describe the resulting superconducting states at the mean field level in terms of a Bogoliubov-de Gennes (BdG) Hamiltonian that contains the basic ingredients of these systems which are:
(1) spin $1\over2$ fermions hopping on 2 and 3D lattices,
(2) mean field order parameters (gap functions) describing pairing.  For chiral superconductivity, this pairing will be intersite.
(3) multiple orbitals per site, which may be split by crystal fields, that form bands crossing the Fermi surface, and
(4) spin-orbit coupling.
The systems that interest us involve all of these ingredients.  \sro\ is a tetragonal layered material with three-bands crossing the Fermi energy and weak to intermediate spin-orbit coupling.\cite{soc1,soc2} \upt\ is  a hexagonal 3D system with six bands crossing the Fermi energy and strong spin-orbit coupling.\cite{abc} In addition to such more realistic models, a great deal can be learned from simpler model systems such as, for example, a single band of spinless or spin $1\over2$ fermions in 2D with nearest neighbor chiral p-wave pairing. Here we discuss the BdG Hamiltonian, incorporating all the basic ingredients descried above, which is the starting point of many calculations of the properties of chiral superconductors.  We also introduce notation and formalism for triplet pairing that will be useful for later sections of the paper.

\subsection{Mean Field Hamiltonians}

We begin by writing the BdG Hamiltonian as ${\cal H}={\cal H}_0 + {\cal H}_{\Delta}$, where 
\begin{equation}
{\cal H}_0=-\sum_{i\ne j}\sum_{a,b,\sigma}t_{i,j}^{a,b}c^{\dagger}_{i,a,\sigma}c_{j,b,\sigma} +
\sum_{i,a,\sigma}(\epsilon_{a}-\mu)c^{\dagger}_{i,a,\sigma}c_{i,a,\sigma}+
\lambda\sum_{i,a,b}\sum_{\sigma,\sigma '}{\vec L}_{a,b}\cdot{\vec\tau}_{\sigma,\sigma '}c^{\dagger}_{i,a,\sigma}c_{i,b,\sigma '}.
\label{h0}
\end{equation}
In this expression, the $-t_{i,j}^{a,b}$ represent the kinetic energy for hopping between the same or different orbitals on different sites.  $\epsilon_{a}$ is the crystal field energy for orbital $a$, and $\mu$ is the chemical potential.  In the third term, $\lambda$ is the spin-orbit coupling constant, ${\vec L}_{a,b}$ is the matrix element of the angular momentum operator between different orbitals on the same site, and the ${\vec\tau}_{\sigma,\sigma '}$ are Pauli matrices.

 Taking the Fourier transform of Eq.~(\ref{h0}) and diagonalizing ${\cal H}_0$ within the space of spin and orbitals for each wave vector, $\vec k$, one may write
\begin{equation}
\label{h0k}
{\cal H}_0= \sum_{\vec k} \sum_{b,\nu} \epsilon_{b,\nu}(\vec k) c^{\dagger}_{b,\nu}c_{b,\nu}
\end{equation}
where $b$ is a band index and where, as a result of the spin-orbit coupling, $\nu$ is a pseudospin index.  For systems with both  inversion and time-reversal symmetry, $\nu$ labels states in a Kramers doublet and $\epsilon_{b,\nu}$, the quasiparticle energy measured with respect to the chemical potential, is independent of $\nu$.

The pairing part of the Hamiltonian, ${\cal H}_{\Delta}$, can be written in this band-pseudospin basis as
\begin{equation} 
\label{hpair}
{\cal H}_{\Delta} =\sum_{\vec k}\sum_{a,b}\sum_{\mu,\nu}\Delta_{a,b}^{\mu,\nu}(\vec k)c^{\dagger}_{a,\mu}(\vec k)c^{\dagger}_{b,\nu}(-\vec k)+ h.c.
\end{equation}
Equations (\ref{h0}-\ref{hpair}) are widely used as the starting point for theoretical calculations discussed in later sections of this paper.
This form of the pairing Hamiltonian includes both intraband ($a=b$) and interband ($a\ne b$) pairing.  If one is only interested in the low energy properties of the superconducting state, one can ignore interband pairing provided $\Delta_0 << E_F$.  In Sec. IV we discuss the Kerr effect at optical frequencies where interband pairing effects can be important.  

Equation (\ref{hpair}) applies equally well to singlet or triplet pairing. However, for the systems we consider, the two are mutually exclusive.   Denoting the pseudospin values by $+$ and $-$, and suppressing $\vec k$ and the band indices, one can write the matrix, $\hat\Delta$, in terms of an s-wave gap and a d-vector that describes the triplet order parameter:\cite{leggett,leggettbook}
\begin{equation}
\hat\Delta = \begin{pmatrix} \Delta^{++} & \Delta^{+-} \\ \Delta^{-+} & \Delta^{--} \end{pmatrix}
=
\begin{pmatrix} -d_x+id_y & d_z+\Delta_s \\ d_z-\Delta_s & d_x+id_y \end{pmatrix},
\end{equation}
where
\begin{align}
\Delta_s&= \frac{1}{2}\bigl(\Delta^{+-} - \Delta^{-+}\bigr) \\
\vec d\ &=\frac{1}{2}\bigl(\Delta^{--}-\Delta^{++},\ 
-i(\Delta^{--}+\Delta^{++}),\ 
\Delta^{+-}+\Delta^{-+}\bigr).
\label{trip}
\end{align}
The singlet case corresponds to $\vec d=0$ and the triplet has $\Delta_s=0$.

Equation (\ref{trip}) describes a completely general spin, or pseudospin, triplet superconductor order parameter.  Here we will be interested in unitary states that satisfy ${\vec d}\times{\vec d}^*=0$. In this case, $\vec d$ is a real vector apart from a $\vec k$ dependent phase factor and the direction of $\vec d$ defines the direction along which the Cooper pair state has zero spin projection. For example, the simple chiral p-wave state, $\vec d = \Delta_0{\hat z} (k_x+ik_y)/k_F$\ is an $S_z=0$\ state.  This can equally well be described as a state with equal gap amplitude for pairing $\uparrow\uparrow$\ and $\downarrow\downarrow$\ (or $++$\ and $- -$) if the spin quantization axis is chosen in the $xy$\ plane.  Such a state is an equal spin pairing state (ESP), defined as a triplet state where the d-vector has no $z$-component for some choice of quantization axis. Another state that will be discussed in later sections is the helical state, $\vec d \sim k_x{\hat x} + k_y{\hat y}$. It follows from Eq.~(\ref{trip}) that this is a state where the up ($+$) spins are in a state with positive chirality, $k_x+ik_y$, and the down ($-$) spins have negative chirality, $-(k_x-ik_y)$. This is another ESP state.  Like chiral p-wave, it is a topological superconducting state, but it has no net chirality and does not break time-reversal symmetry.\cite{chiraltop} 

Non-unitary states have nonzero $|{\vec d}\times{\vec d}^*|$ and non-zero spin polarization, and so they are usually not energetically favourable in zero external magnetic field. In superfluid \he\ a non-unitary state, the A1 phase, is observed only in the presence of an applied magnetic field.\cite{leggett,wheat}

\subsection{Simplified Models}
\label{simplemodel}
The most distinctive features of chiral superconductivity arise even in a simple one-band model.  In that case spin-orbit coupling can be ignored, and, for the triplet ESP case, the spin quantization axes can be chosen so that $\Delta^{\uparrow\downarrow}(i,j)=0$, in which case the problem breaks up into two independent Hamiltonians for $\uparrow\uparrow$ and $\downarrow\downarrow$ Cooper pairs, each of which is effectively a system of spinless fermions.  
Then Eqs. (\ref{h0k}) and (\ref{hpair}) reduce to
\begin{equation}
{\cal H} = \sum_{\vec k ,\sigma}\bigl(\epsilon (\vec k)-\mu\bigr)c^{\dagger}_{\vec k,\sigma} c_{\vec k,\sigma}+\sum_{\vec k ,\sigma}\bigl[\Delta (\vec k)c^{\dagger}_{\vec k,\sigma} c^{\dagger}_{-\vec k,\sigma}+ h.c.\bigr]
\end{equation}
where, for the case of chiral p-wave on a square lattice with nearest neighbor hopping and pairing, 
\begin{align}
\epsilon (\vec k)&=-2t\bigl(\cos(k_x)+\cos(k_y)\bigr) \label{squaree}\\
\Delta (\vec k)&=\Delta_0 \bigl(\sin(k_x)\pm i\sin(k_y)\bigr).
\label{squared}
\end{align}
The quasiparticle excitation energies are given by
\begin{equation}
\label{pbc}
E(\vec k)=\pm\sqrt{(\epsilon (\vec k)-\mu)^2 + |\Delta (\vec k)|^2}
\end{equation}
which is fully gapped around the Fermi surface. In more realistic models discussed later, Eqs.~(\ref{squaree}) and (\ref{squared}) will include further neighbor hopping and pairing.  However, except for the  possibility of accidental zeros on the Fermi surface, $E(\vec k)$ will remain gapped, although it may be highly anisotropic.

\section{Properties of Model Chiral Superconductors}
\label{sec3}
\subsection{P-Wave Chain}
Interesting effects arise when edges and singularities are introduced.    We consider the simplest example, a chain of $N$ sites with open boundaries, which we call the $p$-wave chain and which illustrates how zero energy (Majorana) states result from p-wave pairing. The Hamiltonian for this model is,
\begin{equation}
{\cal H}^x = {\cal H}_t^x +{\cal H}_{\Delta}^x
\end{equation}
where
\begin{equation}
{\cal H}{_t}^x =
\sum_{n=1}^{N-1}\Bigl[-t\bigl( c^{\dagger}_{n+1}c_n +c^{\dagger}_n c_{n+1}\bigr) - \mu c^{\dagger}_n c_n\Bigr]
\end{equation}
and
\begin{equation}
{\cal H}_{\Delta}^x=
\sum_{n=1}^{N-1} \Delta c^{\dagger}_{n+1}c^{\dagger}_n + h.~c.
\label{pairing}
\end{equation}
This model was discussed in the seminal paper of Kitaev,\cite{kitaev} who analysed it in terms of Majorana operators. Here we will continue using (spinless) fermion operators and notation similar to that used in Ref.~(\onlinecite{gurarie}) for the 2D case.

The BdG Hamiltonian  can be written as a  $2N\times 2N$ matrix:
\begin{equation}
h=\begin{pmatrix} \tilde h_t & \tilde\Delta \\ \tilde \Delta^T & -\tilde h_t \end{pmatrix},
\end{equation}
where $\tilde h_t$ is the coefficient matrix of ${\cal H}{_t}^x$\  and $\tilde \Delta$ is antisymmetric so that $\tilde \Delta^T =-\tilde\Delta^*$.
The fact that $\tilde\Delta$ is antisymmetric results from the p-wave property that pairing to the left has the opposite sign as pairing to the right, or, equivalently, from the anticommutation of the fermion operators in the pairing terms, Eq.(\ref{pairing}).

If $\psi =(u,v)^T$ is an eigenfunction of $h$ with energy $E$, it is easy to see that the BdG Hamiltonian, $h$ has the property that $\sigma_x \psi^*=(v^*,u^*)^T$\ is also an eigenfunction with energy $-E$, where 
\begin{equation}
\sigma_x = \begin{pmatrix} 0 & I_N \\ I_N & 0 \end{pmatrix},
\end{equation}
and $I_N$ is the $N\times N$\ unit matrix.  The effect of the operator, $\sigma_x K$, where $K$ is complex conjugation, is to turn a particle with energy $E$ into its anti-particle with energy $-E$.

\begin{figure}
\includegraphics[width=0.6\textwidth]{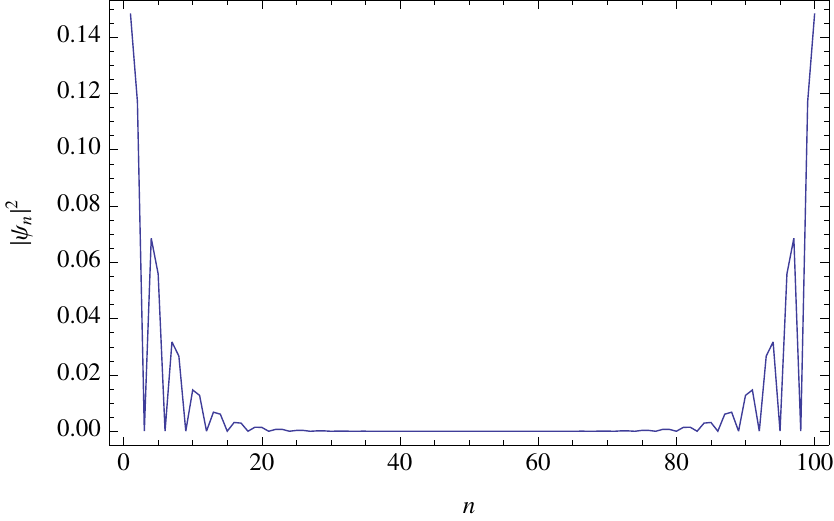}
\caption{\label{edge} Probability density of the near-zero-energy states, $|\psi_n|^2 = |u(n)|^2+|v(n)|^2$, as a function of site index, $n$, for a 1D chain of length 100 for $\Delta = 0.125t$ and $\mu = -t$. As discussed below, these amplitudes are the same for the two states.  Both have their amplitude localized within a coherence length at the two ends of the chain. Even and odd linear combinations of these nearly degenerate states are Majorana modes, $\psi_L$ and $\psi_R$, with amplitude near one end of the chain or the other as discussed in the text.}
\end{figure}

The spectrum of this chain is easily calculated numerically.  For 
$|\mu|<2t$\ and $N$ reasonably large, it is essentially identical to the analog of Eq.(\ref{pbc}) for a 1D ring
with one important difference.  The ring has a gap around zero energy of size $2|\Delta \sin(k_F)|$ while the chain has two states at essentially zero energy, at the center of the gap.  The wave functions of the two ``near-zero modes" are
linear combinations of states localized near the two ends, with a localization length that scales like $t/\Delta$ and energies $\pm E$ where $E$ goes to zero exponentially as $N\Delta/t$ becomes large.  An example, the result of a simple numerical calculation, is shown in Fig.~\ref{edge} where $|u|^2+|v|^2$\ for one of the near-zero energy states is plotted for a chain of 100 sites with $\Delta = 0.125t$ and $\mu = -t$. Similar results appear in the literature for closely related 1D models.\cite{niu}

Say that $\psi_0=(u_0,v_0)^T$ is one of the near-zero modes and $\sigma_x K\psi_0=(v_0^*,u_0^*)^T$\  is the other. Then $\psi_L = (\psi_0 +\sigma_x K\psi_0)/\sqrt{2}$ is localized at one end of the sample and $\psi_R = i(\psi_0 - K\sigma_x \psi_0)/\sqrt{2}$ is localized on the other.   Of course, $\psi_L$ and $\psi_R$ are only energy eigenstates (with $E=0$) in the limit of large N.  Assuming that N is large, we note that these states have the property that each is its own ``antiparticle'' in the sense that $\sigma_x K\psi_L = \psi_L$ and  $\sigma_x K\psi_R = \Psi_R$.  Given the easily satisfied condition that they are well separated spatially, these are true charge-neutral Majorana zero modes whose properties are topologically protected by the existence of the bulk gap.  

\subsection {Edge States and Edge Currents}

\begin{figure}
\includegraphics[width=0.7\textwidth]{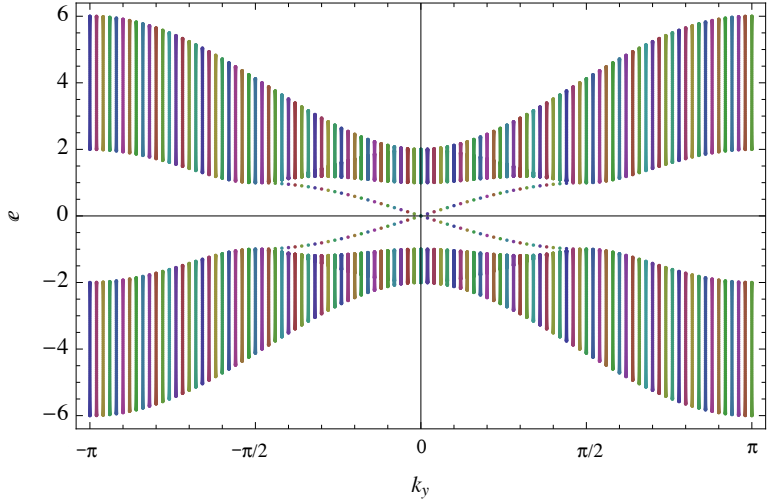}
\caption{\label{2dstrip} Quasiparticle energies, obtained by diagonalizing the BdG Hamiltonian of a chiral p-wave superconductor on an $100\times 100$ square lattice strip with periodic boundary conditions in the $y$-direction, plotted versus $k_y$.  The vertical bars are actually 100 points each for positive and negative energies, except near the center where two modes split off and disperse across the gap. The two branches correspond to Majorana-Weyl states localized within a coherence length on opposite edges of the cylinder. Whether the positive slope branch is localized at one edge or the other depends on the chirality.  The parameters are $\Delta=t$, $\mu = -2t$.}
\end{figure}

For 2D systems with edges, the Majorana modes at the chain ends  hybridize into bands. For the fully gapped 2D chiral p-wave state, these Majorana bands disperse within the gap.
The extension of the above calculation to the 2D chiral case is straightforward.  One diagonalizes the BdG Hamiltonian for an $N_x \times N_y$ chiral p-wave strip with open boundary conditions along $\hat x$ and periodic boundary conditions along $\hat y$.  In this case the spectrum has a continuum of states above and below the gap, and two branches crossing the gap with opposite slopes. A typical numerical result is shown in the Fig.~\ref{2dstrip} for the simple model of Sec.~\ref{simplemodel}.
The two branches correspond to states localized within a coherence length on opposite edges of the cylinder. Whether the positive slope branch is localized at one edge or the other depends on the chirality.  These chiral edge mode are Majorana-Weyl modes,\cite{stone} and are a signature of the topology of the chiral p-wave state.\cite{read}  For chiral superconductors with larger anisotropy or higher chirality (d or f)  the excitation spectra are similar to Fig.~(\ref{2dstrip}), but with additional chiral edge modes in the case of higher chirality.\cite{nontopo,higherchirality}

One can also solve this problem analytically. For example, the continuum limit for one edge, a half-infinite ($ x < 0$) chiral p-wave superconductor, was analyzed by Stone and Roy.\cite{stone} The chiral edge mode for each spin component has the simple dispersion
\begin{equation}
E(k_y)=\mp \Delta_0 k_y/k_F ,
\label{pmode}
\end{equation}
where $k_F$\ is the Fermi wave vector and the sign $\mp$\ is opposite to the chirality of the gap function. For positive chirality, the edge modes with $k_y>0$\ are occupied in the ground state and contribute to a spontaneous equilibrium edge current. The extended states also contribute to this current, cancelling exactly half the contribution from the edge states. Neglecting screening, the total current, which is localized to a coherence length from the edge, is given by\cite{stone}
\begin{equation}
I_y^0=\int_0^\infty  j_y dx = \frac{eE_F}{4\pi}
\label{j0}
\end{equation}
where $E_F=k_F^2/2m$ in the continuum limit and the chirality is chosen as positive. This result is consistent with the prediction of a macroscopic angular momentum of $N\hbar /2$, where $N$ is the total number of electrons.\cite{stone,kita,volovik1}. Eq.~(\ref{j0}) can be written in terms of the density, $\rho$\ rather than $E_F$ to emphasize its coincidence with the topological result expected for a slowly varying density gradient (rather than an abrupt, hard edge boundary condition).\cite{read,nontopo} For a ``soft wall'' where the density falls to zero very slowly, one finds the total current is $I_0=Ce\rho/2m$,\cite{read} which coincides with Eq.~(\ref{j0}) if the Chern number, $C$, is +1. The Chern number is a topological number, an integer given by the winding of the chiral order parameter around the Fermi surface, and is defined as
\begin{equation}
C=\frac{1}{4\pi}\int d{\bf k}\ {\hat h}\cdot (\partial_{k_x}{\hat h}\times\partial_{k_y}{\hat h}),
\label{chern}
\end{equation}
where ${\bf h}=\bigl({\rm Im}[\Delta({\bf k})],{\rm Re}[\Delta({\bf k})],\epsilon(k)-\mu \bigr)$. The sign of the Chern number depends on the chirality and the sign of the charge carriers. In the case of multiple bands, it is defined for each band and the total Chern number is the sum. The simple chiral p-wave case, discussed above and shown in Fig.~(\ref{2dstrip}) has Chern number $C=\pm 1$.

Eq.~(\ref{j0}) can be generalized for an arbitrary band structure in two or three dimensions, using the quasiclassical approximation,\cite{nontopo}
\begin{equation}
\label{Igen}
I_y=\frac{e}{(2\pi)^d}\oint_{FS} \frac{d\textbf{p}}{|{\textbf v}|}v_xv_y\tan^{-1}\Bigl( \frac{\Delta_x}{\Delta_y}\Bigr).
\end{equation}
Here $\vec v=\partial \epsilon /\partial \vec k$\ and the chiral gap function is $\Delta_x +i \Delta_y$.
For the case of a square lattice with only nearest neighbour hopping and pairing, i.e. the model in Sec.~\ref{simplemodel}, this integral gives the same result as Eq.~(\ref{j0}) for any filling of the band if one interprets $E_F$\ as the chemical potential measured from the bottom of the band for $E_F<0$\ and measured from the top of the band (as is appropriate for holes) for $E_F>0$.\cite{nontopo}  The results of numerical BdG for the simple square lattice model also coincide with this analytic result, even for self-consistent BdG where one finds $\Delta_x$\ is suppressed within a coherence length of the surface, an effect which is ignored in the quasiclassical approximation.\cite{nontopo}  This may seem surprising, since the edge current itself is not topologically protected and can depend on microscopic details. However, this coincidence can be understood,  using spectral flow arguments which apply because of the separability of the $x$\ and $y$\ degrees of freedom in the nearest neighbour model, as resulting from a combination of the topology and symmetry which require a single chiral edge mode passing through zero energy at $k_y=0$.\cite{nontopo} For a more realistic bandstructure that includes second neighbour hopping, the current is still given by Eq.~(\ref{Igen}) but is reduced from that of Eq.~(\ref{j0}).

For a hexagonal lattice structure, chiral d-wave may be stable. In the continuum limit, the chiral d-wave gap function can be written as $\Delta_0[(k_x\pm ik_y)^2]/k_F^2$. The Chern number in this case is $\pm$2 and there are two chiral edge modes at each edge. In this case, the chiral edge mode dispersion, given by Eq.~(\ref{pmode}) for the p-wave case, becomes\cite{higherchirality}
\begin{align}
E(k_y) &= \mp\Delta_0(k_F^2-2k_y^2)/k_F^2\qquad  & -k_F<k_y<0 \\ 
  &=\pm \Delta_0(k_F^2-2k_y^2)/k_F^2\qquad & 0<k_y<k_F . 
\end{align}
In contrast to the chiral p-wave, the edge current vanishes in this case (and, in fact, vanishes for any higher angular momentum chiral superconducting state in the continuum limit) although it recovers the full topological value of $eE_F/2\pi$, corresponding to a macroscopic angular momentum of $N\hbar$, for a ``soft wall''.\cite{higherchirality}  While the presence of two zero energy modes per edge is protected by topology, the position of these zeros with $k_y$\ is not protected and there can be spectral flow.\cite{oshikawa,nontopo} Once lattice effects are included, the current can be non-zero but is significantly reduced from the chiral p-wave case.\cite{higherchirality} This may be relevant to \upt\ and other materials thought to be chiral d- or f-wave.

So far the discussion has focussed on simple chiral superconductor models. For more realistic models, the current can differ significantly from Eq.~(\ref{j0}), even changing sign if there are extra zeros in the chiral edge modes that allow for spectral flow\cite{nontopo}. Such extra zeros can result from large gap anisotropy caused by large next-nearest or further neighbor pairing\cite{nontopo,largechern} or from retroreflecting\cite{retro} or facetted\cite{facet} edges. Combining these effects with strong disorder in the edge region can further suppress the current.\cite{lederer,largechern}

Screening has been neglected in the above discussions. The effect of screening on the chiral p-wave edge current has been calculated in the quasi-classical limit of the continuum model.\cite{matsumoto} For parameters appropriate for \sro\ the maximum magnitude of the local current near the edge is reduced by about a factor of 10 and would give rise to a magnetic field near the edge of the order of 10G. The maximum field at an idealized internal domain wall is ~20G.\cite{matsumoto} These are the results to which experiments on \sro, discussed in Sec.~\ref{secSRO}, are compared. 

\subsection{Vortices}

The case of a vortex in a $p_x\pm ip_y$ superconductor and its zero energy Majorana core state has been worked out in detail by Tewari et al.\cite{tewari}.  For a spinless chiral p-wave vortex, there is a single zero energy mode that is a Majorana mode and robust to perturbations.\cite{tewari}  For the usual spinful case, there are two zero-modes corresponding to the two quasiparticle spin states. Due to this degeneracy, these zero energy states are not as robust as they are in the spinless case. For example, Zeeman coupling shifts these zeros to positive and negative energies for the two spin states.\cite{bauer}  However, the spinful case can support half-quantum vortices (HQVs) with a robust zero energy Majorana mode.

To understand HQVs it is useful to note that an  $S_z=0$ p-wave gap function, $\Delta({\bf k}){\hat z}$, could equally well be written as 
\begin{equation}
e^{i\theta}\Delta({\bf k})\bigl[-e^{-i\phi} |\uparrow\uparrow\rangle +e^{i\phi} |\downarrow\downarrow\rangle\bigr],
\label{hqv1}
\end{equation}
where  $\theta$ is the U(1) phase and the spin quantization axis is chosen in the $xy$-plane, with angle $\phi$\ from the $x$-axis. Equivalently, if we fix the quantization axis along $x$ ($\phi$=0), then changing $\phi$ is equivalent to rotating the d-vector. A regular vortex corresponds to a state where the U(1) phase $\theta$ changes by 2$\pi$, with no change in the d-vector. A HQV corresponds to a state where $\theta$\ changes by $\pi$, and $\phi$\ also changes by $\pi$. It follows from Eq.~(\ref{hqv1})  that the gap function is single-valued, with no net winding in the up-spin component and a 2$\pi$\ winding in the down-spin component. Since only the charge, not the spin, couples to magnetic flux, the $\pi$\ winding of the U(1) phase corresponds to half the usual superconducting flux quantum, or $\Phi_0/4e$, where $\Phi_0$\ is the fundamental flux quantum.  It also follows from Eq.~(\ref{hqv1})\ that the d-vector rotates from +$z$\ to -$z$\ in going around the HQV. 

In addition to the usual charge supercurrents associated with a regular vortex, which are screened, a HQV, since it is a vortex in only one spin component (or equivalently, since the spin state changes as one goes around the vortex) also has spin currents that are not screened and cost an energy which grows logarithmically with the system size. If one has a very small system size, this may not be a large penalty.\cite{sukbum} A HQV may also be costly in energy if the d-vector is strongly pinned in one-direction due to spin-orbit coupling. 

Note that nowhere in the above discussion of HQVs was $\Delta({\bf k})$\ specified to be chiral. Any ESP state can support HQVs. For a simple chiral (or helical) p-wave state, there is a single robust Majorana zero mode in the HQV core.\cite{tewari} This follows from the fact that only one spin state is involved in the vortex. HQVs will be discussed further in the Section on \sro.

\section{Anomalous Hall and Polar Kerr Effects}
\label{secKerr}

Broken time reversal symmetry (BTRS) is a defining property of chiral superconductivity and a number of probes are used to detect BTRS in superconductors including muon spin resonance, Josephson tunneling, scanning SQUID or Hall bar probes, and polar Kerr measurements.  These will be discussed in the following sections on \sro\ and \upt. However, the polar Kerr effect, which is directly related to the anomalous Hall effect, deserves further discussion here, as it has caused some confusion in the literature, both in connection to \sro\ as well as to the high \tc\ cuprates, and Kerr effect experiments on these materials have driven some recent theoretical advances in our understanding. See, for example, Refs.~\onlinecite{roy} and \onlinecite{cupratekerr}, which address earlier misconceptions about the Kerr effect.

In a polar Kerr experiment, linearly polarized light of frequency, $\omega$, normally incident on the sample, is reflected as elliptically polarized light with the polarization axis rotated by the Kerr angle, $\theta_K$, which is related to the Hall conductivity:
\begin{equation} 
\label{kerrangle}
\theta_K(\omega) = \frac{4\pi}{\omega}\textrm{Im}\Bigl(\frac {\sigma_H(\omega)}{n(n^2-1)} \Bigr) , 
\end{equation}
where $n$ is the complex index of refraction and $\sigma_H(\omega)=[\sigma_{xy}(\omega)-\sigma_{yx}(\omega)]/2$ is the Hall conductivity.  It follows from Eq.~(\ref{kerrangle}), for general $\omega$, that one has a non-zero Kerr angle if and only if the Hall conductivity is non-zero. Thus, a polar Kerr effect implies an anomalous Hall effect (a Hall effect in the absence of an applied magnetic field).  While the index of refraction can introduce strong frequency dependence, particularly for frequencies close to the plasma frequency, it suffices to study $\sigma_H(\omega)$ to understand how a chiral superconductor gives rise to a non-zero Kerr effect.

BTRS is a necessary but not sufficient condition to have a Kerr effect or an anomalous Hall effect in a chiral superconductor. Broken translational symmetry is also necessary because the external field only couples to the center-of-mass momentum which is decoupled from the relative degrees of freedom (i.e., interaction effects) in a Galilean invariant system.\cite{read} The BTRS of a translationally invariant chiral superconductor could, in principle, be probed at finite wave vector. For a clean, isotropic two-dimensional chiral p-wave superconductor at zero temperature, one can show the $q$-dependent anomalous Hall conductivity is\cite{roy,goryo99,golub}
\begin{equation}
\sigma_H(q,\omega)=\frac{e^2C}{2\hbar}\frac{v_F q^2}{\omega^2- v_F q^2} , 
\end{equation}
where $v_F$ is the Fermi velocity, C is the Chern number and $\omega<<\Delta_0$. At high frequencies, this expression is suppressed by a factor of $(\Delta_0/\omega)^2$.\cite{roy} Note that $\sigma_H(0,\omega)$, vanishes, as expected for a translationally invariant chiral superconductor.  While it would be interesting to probe the $q$-dependent Hall conductivity in the limit where $v_F q$ is comparable to or larger than $\omega$, it is not clear how one could do this. The finite beam size of the incident photons in a polar Kerr experiment does bring in a finite in-plane ${\bf q}$, but this effect is too small to be detected.\cite{roy}

Impurities in a chiral superconductor give rise to a Kerr effect, or $\sigma_H$, although the lowest order Born contribution, of order $n_IU^2$, where $n_I$ is the density of impurities and $U$ their potential, vanishes. The dominant impurity contribution for chiral p-wave is from a type of ``skew scattering'' and is of order $n_iU^3$.\cite{goryo,yakovenko,spivak} This contribution vanishes for higher angular momentum pairing, such as chiral d- or f-wave, in the continuum limit.\cite{goryo} Presumably, order $(n_IU^2)^2$ or higher terms would contribute for higher angular momentum pairing.

To date, the anomalous Hall conductivity of possible chiral superconductors has only been measured at high frequencies, a substantial fraction of an eV, through measurements of the Kerr effect.  At high frequency the single particle contribution to the anomalous Hall conductivity dominates and $\sigma_{xy}$ can be written as:\cite{taylor}
\begin{equation}
\label{sigmaxy} 
\sigma_{xy}(\nu_n)=\frac{ie^2T}{\nu_n}\sum_{{\bf k},\omega_n} Tr[v_x G_0({\bf k},\omega_n)v_y G_0({\bf k},\omega_n+\nu_n)], 
\end{equation}
where $G_0$ is the Greens function in the Nambu representation, $\nu_n,\omega_n$ are Bose and Fermi Matsubura frequencies and $v_i$ is the $i$-component of the velocity matrix. For a single band chiral superconductor $v_i=\partial\epsilon(k)/\partial k_i$, multiplying the 2x2 identity matrix. In that case, $v_i$ commutes with $G_0$ and it follows that $\sigma_{xy}=\sigma_{yx}$. In other words, at the single particle level, $\sigma_H$ vanishes for any one-band chiral superconductor.

There are higher order contributions (i.e., vertex corrections) to the anomalous Hall conductivity of a clean one-band chiral superconductor. Yip and Sauls identified a contribution that brings in the flapping mode (oscillations in the direction of the angular momentum vector) for a 3-dimensional chiral p-wave superconductor.\cite{yip} In this case, the translational symmetry is broken perpendicular to the surface (whose normal is taken to be parallel to the angular momentum vector of the chiral superconductor). The effect is non-zero if one accounts for the incident photon momentum and the finite coherence length along the photon momentum, and if particle-hole symmetry is broken. This brings in three small factors for a quasi-2d chiral superconductor. However, in a clean chiral superconductor, this contribution, while small, may dominate at frequencies comparable to the gap. In a multi-band chiral superconductor another contribution has been identified which is noticeably larger at frequencies well above the gap.\cite{taylor,taylor2,annett1,annett2}

\begin{figure}
\includegraphics[width=0.6\textwidth]{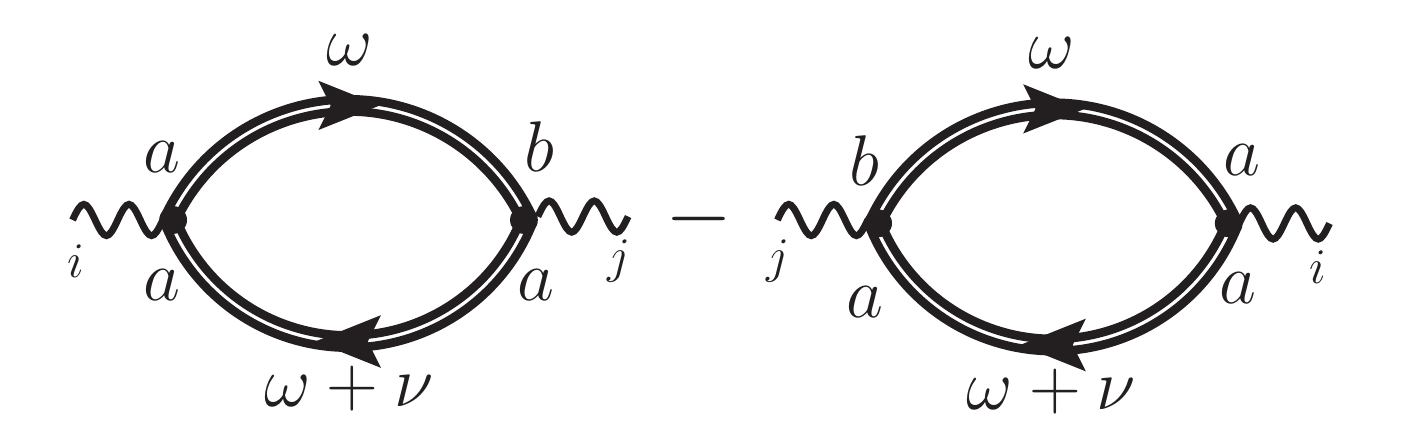}
\caption{\label{diagram} One-loop contribution to the intrinsic anomalous Hall conductance for a multi-band chiral superconductor, where $a\ne b$ are orbital indices and $(i,j)=(x,y)$ denote the photon polarization. At one vertex, a photon of frequency $\nu$\ induces an interorbital transition. }
\end{figure}
For a multi-band or multi-orbital chiral superconductor, Eq.~\eqref{sigmaxy} still applies but now $G_0$ and $v_i$ are, for example, 4x4 matrices in the case of 2 bands. Importantly, $v_i$ is no longer simply proportional to the identity matrix, allowing for the possibility of a non-zero intrinsic Hall conductivity.  Indeed, there is one type of contribution to $\sigma_{xy}$ that is not fully symmetric under interchange of $x$ and $y$.  This contribution is shown diagrammatically in Fig.~\ref{diagram}, where $a\ne b$ are orbital labels.\cite{taylor}  In the band basis, such a contribution requires interband pairing.\cite{taylor}  The contribution shown in Fig. \ref{diagram} to the imaginary part of $\sigma_H$, which is the absorptive part, vanishes at zero temperature for frequencies below an electronic energy scale, the smallest energy difference between states on the two different bands at the same momentum. For \sro, this electronic energy scale is $t''$, the hopping matrix element between $xz$ and $yz$ orbitals on neighbouring sites.\cite{taylor2} By contrast, the impurity and flapping mode contributions to Im($\sigma_H$), discussed above, are nonzero for frequencies above $2\Delta_0$.  However, at high frequencies, well above the gap energy, the multi-band contribution is likely to be the dominant contribution to $\sigma_H$ in a multi-band chiral superconductor in the clean limit.

One should be able to distinguish between an intrinsic and an extrinsic (due to impurities) Kerr effect or anomalous Hall effect by comparing measurements on samples with varying amounts of disorder.  If one is in the clean limit, $\Delta_0$ is not strongly impurity dependent for small changes in impurity concentration, while the predicted $\sigma_H$ changes linearly with impurity concentration. In addition, measurements of the Kerr effect or Hall conductivity at multiple frequencies, should be able to distinguish between different intrinsic effects, such as the multi-band and collective mode effects discussed here.

\section{S\lowercase{r}$_2$R\lowercase{u}O$_4$\ and Chiral P-Wave Superconductivity}
\label{secSRO}

We now turn our attention to real materials that are thought to exhibit chiral superconductivity, starting with \sro.
Superconductivity was discovered in \sro\ in 1994 by Yoshi Maeno and collaborators,\cite{maeno94} and shortly after theoretical work pointed to chiral p-wave order based on analogies to superfluid He-3.\cite{rice95,baskaran} Since then many different experiments give evidence for triplet, odd-parity pairing and chiral order, making \sro\ one of the strongest candidates for chiral superconductivity, despite some puzzles remaining.\cite{puzzles}  Here, the key properties of \sro\ are very briefly reviewed, some of the experiments and theoretical work addressing the nature of the superconductivity are highlighted, and some outstanding questions are discussed. More comprehensive reviews include Refs.~\onlinecite{maenotriplet}, \onlinecite{Mackenzie} and \onlinecite{kallin}.

\subsection{Key Properties of \sro}

\begin{figure}
\includegraphics[width=0.4\textwidth]{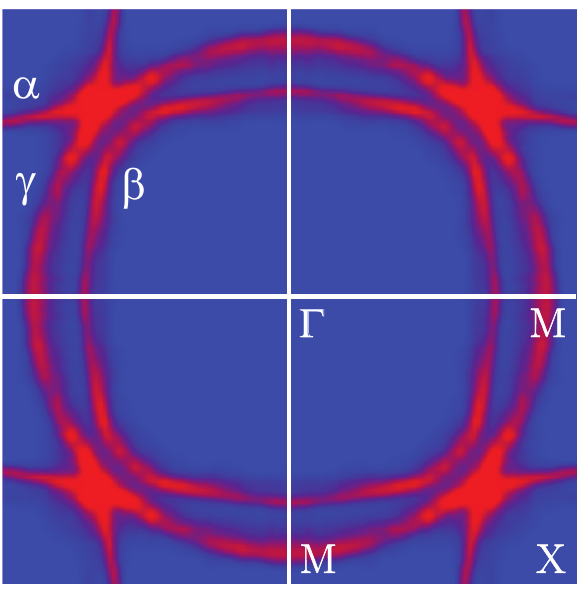}
\caption{\label{sroARPES} Fermi surfaces of \sro\ as determined by ARPES by Damascelli et al.\cite{damascelli}. The hole band, $\alpha$, and the electron band, $\beta$, result from hybridization of the quasi-one-dimensional Ru $d_{xz}$ and $d_{yz}$ bands.  The electron band, $\gamma$, is primarily formed from Ru $d_{xy}$ orbitals.}
\end{figure}

\sro\ has a tetragonal crystal structure similar to that of the cuprate superconductors and, like the cuprates, is a highly anisotropic layered material, with conducting RuO$_2$ layers (the $ab$ planes, also taken to be $xy$ here) and much weaker conduction along the $c$ (or $z$) axis. \sro\ behaves like a Fermi liquid below about 50K, although with substantial mass and susceptibility enhancements, reflecting the strong electronic correlations.\cite{Mackenzie} Three bands, derived from the Ru $t_{2g}$\ 4$d$-orbitals (hybridized with oxygen orbitals), cross the Fermi energy. The $\gamma$\ band, derived primarily from $d_{xy}$\ orbitals, is electron-like. The $d_{xz}$\ and $d_{yz}$\ orbitals form one-dimensional bands that mix to form the hole-like $\alpha$\ and electron-like $\beta$\ bands.  The Fermi surface, in the ab plane, obtained by ARPES\cite{damascelli}, is shown in Fig.~\ref{sroARPES}. The 3 Fermi sheets, particularly the $\gamma$\ sheet, have very little dispersion along the $c$-axis. If one neglects the very small interlayer hopping, then the $d_{xy}$ orbitals only mix with the  $d_{xz}$\ or $d_{yz}$\ through spin-orbit coupling (SOC). Estimates of the SOC vary from about 40 meV to 100meV or larger,\cite{soc,soc1,soc2} with spin-ARPES experiments giving a value of 130 meV at the Gamma point,\cite{spinarpes} while the largest in-plane hopping parameter is estimated to be 250-400meV. 

There are additional properties of the normal state that are relevant for the stabilization of unconventional (non-s-wave) superconductivity. No magnetic order is seen in \sro, but both ferromagnetic and incommensurate antiferromagnetic fluctuations are present in the low-temperature normal state.\cite{Mackenzie} The incommensurate antiferromagnetic fluctuations are associated with the approximate nesting of the $\alpha$ and $\beta$\ bands. In addition, since the 3 bands are 2/3 filled (i.e., 4 electrons/site), the Fermi surfaces in the ab plane are large and the $\gamma$\ sheet is in fairly close proximity to the van Hove point at the Brillouin zone boundary. (See Fig.~\ref{sroARPES}.)

Superconductivity is observed below 1.5K in the best crystals and is strongly disorder dependent as expected for unconventional (non-s-wave) pairing.\cite{disorder} Early experiments (see below) found evidence for triplet pairing with BTRS.  If one takes these experiments at face value and assumes p-wave pairing with broken time-reversal symmetry,  then the tetragonal crystal structure with a cylindrical Fermi surface, picks out chiral p-wave pairing with the symmetry of $(k_x\pm i k_y){\hat z}$. The $k_x$, for example, may be a linear combination of sin$k_x$, sin$k_x$cos$k_y$, sin3$k_x$, etc., depending on whether first, second, or further neighbor pairing is appreciable. Some energetic arguments go into uniquely picking out this pairing symmetry. First, f-wave is also possible, depending on the microscopic details, but theory for \sro\  (discussed below) finds either p- or d-wave is most stable. Second, there are other p-wave states with BTRS, the non-unitary states defined by non-zero ${\bf d}\times {\bf d^*}$.\cite{domain2} As discussed earlier, these states are usually not energetically favored in zero external field. 

\subsection{Microscopic Theory of Superconductivity in \sro}

Developing a microscopic theory of superconductivity in \sro\ is an ongoing endeavour\cite{Mackenzie,maenotriplet} and here we simply highlight some of the developments in the last few years. Superconductivity in \sro\ is thought to arise from magnetic fluctuations and Coulomb repulsion, although there is evidence that phonons may also play a role.\cite{Mackenzie,phonon} If one includes only on-site interactions, the microscopic, two-dimensional Hamiltonian is 
\begin{equation}
\label{htot}
{\cal H} = {\cal H}_0 + {\cal H}_{int}
\end{equation}
where ${\cal H}_0$ is given by Eq.~\ref{h0}. The on-site interactions are
\begin{eqnarray}
\nonumber
\label{hint}
&{\cal H}_{int}= \frac{U}{2}\sum\limits_{i,a\sigma\ne\sigma '} n_{ia\sigma}n_{ia\sigma '} +
       \frac{U'}{2}\sum\limits_{i,a\ne b,\sigma,\sigma '}  n_{ia\sigma}n_{ib\sigma '} \\
&+
     \frac{J}{2}\sum\limits_{i,a\ne b,\sigma,\sigma '} c^{\dagger}_{ia\sigma}c^{\dagger}_{ib\sigma '}c_{ia\sigma '}c_{ib\sigma}+
   \frac{J}{2}\sum\limits_{i,a\ne b,\sigma\ne\sigma '}c^{\dagger}_{ia\sigma}c^{\dagger}_{ia\sigma '}c_{ib\sigma '}c_{ib\sigma}
\end{eqnarray}
where $n_{ias}=c^{\dagger}_{ia\sigma}c_{ia\sigma}$ and $U'=U-2J$.\cite{dagotto}

This Hamiltonian has been studied within weak-coupling renormalization group (RG), an exact approach for infinitesimal $U$ and $J$. Setting $J=0$\ and neglecting spin-orbit coupling, d-wave superconductivity is favoured on the $\gamma$\ band, but even stronger chiral p-wave superconductivity is favoured on the $\alpha$\ and $\beta$\ bands.\cite{raghu} Including spin-orbit coupling perturbatively, chiral p-wave is stable on all three bands for $J=0$ in the weak-coupling limit.\cite{raghu}
Functional RG studies of Eqs.~(\ref{htot}-\ref{hint}) with non-zero $J$, found similar results with chiral p-wave on all 3 bands, although the $\gamma$ band dominated with a substantially larger gap.\cite{frg} Functional RG is non-perturbative and, consequently, can deal with strong interactions, $U$ and $J$, (although, in practice, the equations are truncated, which introduces errors). The dominance of the $\gamma$\ band is understood as resulting from its close proximity to the van Hove singularity and finite interactions bringing the proximate van Hove points into play.\cite{frg}

Scaffidi, Romers and Simon\cite{scaffidi} studied the Hamiltonian in Eqs.~\ref{htot}-\ref{hint}, including spin-orbit coupling, as a function of $J/U$ within weak-coupling RG and found the chiral state becomes unstable to helical p-wave order at $J/U\approx 0.065$, while d-wave order is found for $J/U > 0.3$. The helical state has the symmetry of ${\bf d}=k_x{\hat{\bf x}}+k_y{\hat{\bf y}}$, the 2D analogue of the B phase of superfluid He-3.\cite{leggett} In the chiral (helical) phase the superconducting gap amplitude is largest (smallest) on the $\gamma$\ band. Close to the transition from chiral to helical (as $J/U$ is varied) the superconducting gap on all three bands is comparable. The relative size of the gap on different bands is still an open question, but fits to specific heat data suggest that the gap is comparable (within a factor of 2 or 3) on all three bands.\cite{davis} This is consistent with weak-coupling RG results for either helical or chiral order for a range of $J/U$ in the vicinity of 0.06.\cite{scaffidi,davis}

One key result from the weak-coupling RG calculations is that, within the approximation, the superconducting order is highly anisotropic on all 3 bands, with particularly deep minima on the $\beta$\ band close to the [110] direction.\cite{,raghu,scaffidi} This results from incommensurate spin fluctuations which strongly favour sin3$k_x$\ and sin$k_x$cos$k_y$\ pairing over nearest-neighbour sin$k_x$ pairing.\cite{scaffidi,ohno,largechern} Large gap anisotropy can affect many of the physical properties of the chiral p-wave state. The deep minima will show up as low-lying excitations down to temperatures of order the minima which, in the weak-coupling calculations, is an order of magnitude smaller than the maximum gap. Also, as mentioned earlier, large gap anisotropy can substantially reduce the edge currents, particularly in the presence of strong surface disorder.\cite{largechern,nontopo} This could reconcile some of the apparent discrepancies between experiments on \sro\ and chiral p-wave order.  Consequently, further experiments that could reveal information about gap anisotropy would be of great interest.

 The above RG analysis uses a small value for the spin-orbit coupling, about 10\% the primary hopping. It would be interesting to see if the weak-coupling and functional RG results change qualitatively with stronger SOC.\cite{puetter} Finally, the RG calculations use 2D models. While this seems reasonable given the large anisotropy, the c-axis coherence length is about 30$\AA$, larger than the interlayer spacing. 3D models have been proposed with either chiral p-wave essentially unchanged from the 2D case or with horizontal nodes, i.e. an additional factor of cos $k_z$\ or sin$k_z$\ in the gap function.\cite{annett,mazin}. Several experiments show clear evidence of low-lying excitations in the superconducting state of \sro, but it is still an open question as to whether these are from horizontal nodes, vertical nodes, or only very deep gap minima.\cite{Mackenzie,kallin}

\subsection{Triplet Pairing and Half-Quantum Vortices}

Experiments that point toward triplet (or odd-parity) pairing include spin susceptibility,\cite{nmr1,nmrc} polarized neutron diffraction,\cite{neutron} Josephson tunneling,\cite{liu} and magnetometry measurements consistent with half-quantum vortices.\cite{budakian} These provide substantial evidence for triplet superconductivity. For example, NMR measurements with an in-plane ($ab$) field see no suppression of the spin susceptibility at low temperatures below \tc; the Knight shift remains constant through and below \tc\ as predicted for the chiral p-wave state.\cite{nmr1,leggett}  Very recent NMR measurements even see a tiny increase in the Knight shift below \tc,\cite{imada-new} an effect first predicted for the A phase of He-3 but not observed.\cite{leggett,theory}  

The Knight shift is also constant with temperature for fields along {\bf c},\cite{nmrc} This is inconsistent with the chiral state which has $\langle S_c\rangle=0$ and, consequently, the spin susceptibility should be suppressed for fields along {\bf c}. 
It has been proposed that the d-vector rotates in a field, as is known to happen in He-3.\cite{leggett} Since the helical state with an in-plane d-vector is expected to be close in energy to the chiral state, a field of 200G may be sufficient to cause a first order transition from a chiral to a helical state.\cite{drotation} Direct evidence of such a transition would be further compelling evidence for triplet pairing.

Of course, one needs to be careful about interpreting Knight shift data in a multi-band superconductor with spin-orbit coupling.  While calculations of the Knight shift for realistic models of \sro\ in the superconducting state have yet to be done, Pavarini and Mazin \cite{Pavarini} point out that in the normal state of \sro\ there can be cancellations between the different bands so that the Knight shift could even increase below \tc\ for a singlet superconductor, depending on the size of the superconducting gap on different bands. Other proposals for the superconducting order, including singlet, d-wave order with the symmetry of $k_z(k_x\pm ik_y)$, also written as $d_{xz}\pm id_{yx}$, are compatible with some experiments.\cite{mazin}  Still, it would be surprising to have multi-band and spin-orbit effects conspire to give a constant or very slightly increasing Knight shift in a singlet state. Furthermore, other experiments, including ones that point toward the existence of half-quantum vortices, appear to be incompatible with singlet pairing.

\begin{figure}
\includegraphics[width=0.6\textwidth]{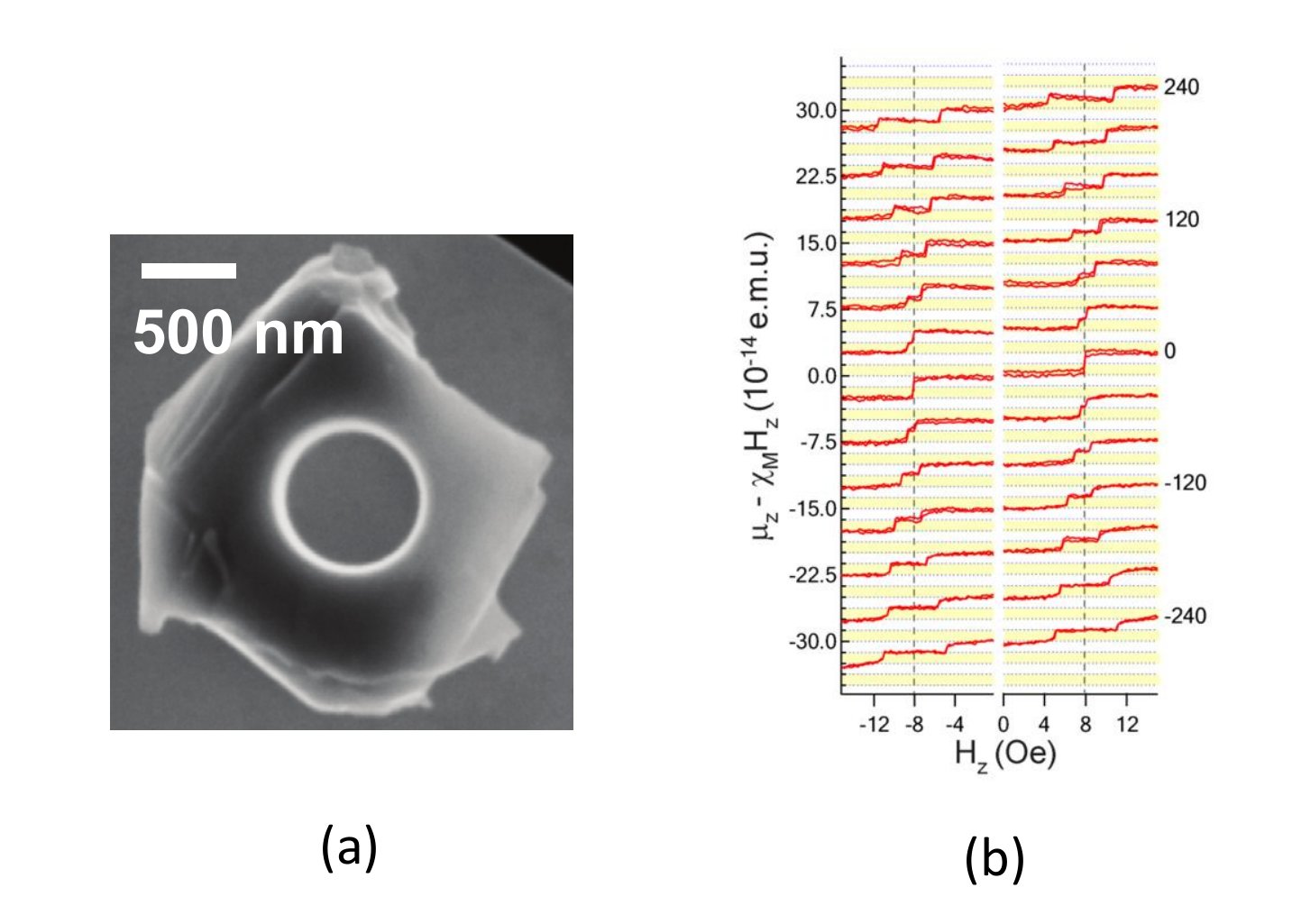}
\caption{\label{hqv} Evidence consistent with half flux quantum vortices in \sro\ from Ref.~\onlinecite{budakian}. (a) An example of the submicron samples of \sro. (b) The magnetization along $\hat z$ is measured as a function of magnetic field along $\hat z$ for different values of an in-plane field H$_{\textrm{x}}$.  The steps seen for $H_{\textrm{x}}=0$ split into two half-steps for $H_{\textrm{x}}\ne 0$ }
\end{figure}

As discussed in Sec.~\ref{sec3}, ESP triplet superconductors can support half-quantum vortices (HQVs).  Budakian and collaborators did ulta-sensitive magnetometry measurements on sub-micron sized samples of \sro\ with a hole milled through the center as shown in Fig.~\ref{hqv}.\cite{budakian} As a field along the c-axis is turned on and increased, screening currents will flow around the milled hole to ensure a quantized fluxoid that steps through integer values of the superconducting flux quantum. The magnetization of the sample will display steps as the fluxoid changes from 0 to 1, 1 to 2, etc., and these steps are observed in Ref.~\onlinecite{budakian}.  When an additional in-plane magnetic field was applied, each of these steps split into two half-steps as shown in Fig.~\ref{hqv}.  
This is consistent with the theory of HQV, if one assumes that the energetics are such that the HQV is not energetically favourable until a field is applied in the ab-plane that couples to the spontaneous magnetization of the HQV, lowering the energy.\cite{roberts}  One might worry about the possibility of wall vortices, that is vortices entering the sample through the $ac$\ or $bc$\ faces and exiting either through the milled hole or through the opposite $ac$\ or $bc$\ face. If these vortices cut through half of the sample (half measured along the $c$-direction) then they would look like a HQV. Ref.~\onlinecite{budakian} took care to try to rule out this possibility. Ideally, one would like to see a complementary technique used to observe and confirm HQV, for example, in a Little-Parks type experiment where the change in \tc, as the flux through the hole of a cylindrical sample is varied, is reflected in measurements of the resistivity close to \tc. The results of such an experiment for \sro\ has been reported and no evidence of HQV was observed.\cite{xcai} However, these results are in the absence of any in-plane magnetic field, where the magnetometry measurements also saw no evidence for HQVs in equilibrium.\cite{budakian} 

\subsection{Evidence for Broken Time Reversal Symmetry and Chiral Superconductivity}

Early evidence of BTRS in \sro\ came from $\mu$SR experiments that see additional muon spin relaxation below \tc.\cite{musr1,musr2}  These experiments are roughly consistent with the fields one would expect from a simple chiral p-wave model for internal domain walls separating the two chiralities, with an average linear domain size of $\sim$10 microns in the $ab$\ layers.\cite{matsumoto} 

The Kerr effect is another key probe of TRSB. A Kerr angle of about 100 nrads is observed in \sro\ at a probing frequency of 0.8eV.\cite{kapitulnik} This might be explained by impurities.\cite{goryo,yakovenko} Alternatively, if it is an intrinsic effect, it is then likely to be due to interband pairing. This has been estimated, with parameters appropriate for \sro, to give rise to a Kerr angle of the same order as the observed angle, provided there is substantial superconductivity on the quasi-1D bands.\cite{taylor,taylor2,annett1,annett2}  More information could be obtained by measuring the Kerr effect or the anomalous Hall effect on purposely disordered samples or at lower frequencies as discussed in Sec.~\ref{secKerr}.

\begin{figure}
\includegraphics[width=0.6\textwidth]{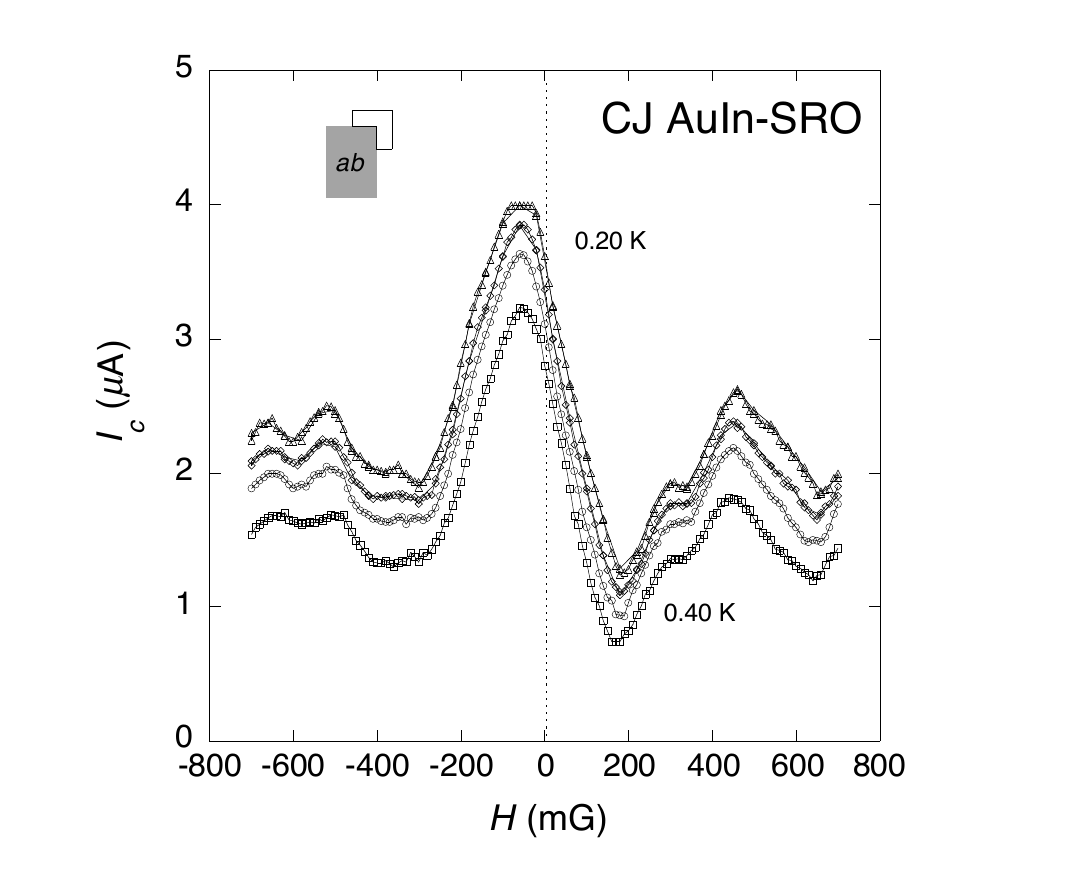}
\caption{\label{corner} Evidence for broken time-reversal symmetry in \sro\ from Ref.\onlinecite{liu}. Quantum interference in I$_{\textrm{c}}$(H) for a corner junction on two adjoining ac and bc faces of Sr2RuO4 at T = 0.20, 0.30, 0.35, and 0.40 K. In this sample, I$_{\textrm{c}}$(H=0) fell roughly half-way between its maximum and minimum values, as is expected for a chiral p-wave superconductor, but that behavior was not reproduced in any other sample.}
\end{figure}

Josephson tunneling experiments can give phase information about the gap function and, consequently, about the BTRS or chirality. Josephson interferometry applied to a corner junction, as shown in the inset of Fig.~\ref{corner} can be used to distinguish between s-wave, d-wave and chiral p-wave gap symmetry. With the two connected junctions on either side of the corner (i.e. on $ac$\ and $bc$\ faces) and a magnetic field applied along the c-axis, an s-wave superconducting sample would display a critical current maximum at zero magnetic flux. A d-wave superconductor would display a minimum at zero flux because of the relative $\pi$ phase shift of the gap function on the two faces and this was observed in the cuprate superconductors.\cite{wollman} For a chiral p-wave superconductor, one expects zero magnetic flux to be roughly half way in-between a maximum and minimum, because of the $\pi/2$ phase shift. This was was seen for \sro\ in one corner junction,\cite{liu} as shown in Fig.~\ref{corner}. 
This effect was not reproduced in any other corner junction, perhaps because of the presence of domain walls. Similar measurements with the junctions applied on opposite faces (e.g. on opposite ac faces) did reproducibly yield the signal expected for a $\pi$ phase shift, compatible with p-wave pairing.\cite{liu} In this case, domains appear not to be a problem. On the other hand, similar measurements on the same face where one would expect to detect no phase shift for any pairing symmetry, or possibly $\pi$ phase shift for chiral p-wave if there were domains, yielded complex Fraunhoefer-like behavior.\cite{dale} It was shown that one could obtain somewhat similar curves with chiral p-wave and domain walls, provided one included band anisotropy and assumed that the domain walls intersected the sample edge at oblique angles.\cite{sigrist-domain}  The conclusion of all these Josephson interferometry measurements on \sro\ is that there are signals of TRSB provided one makes some assumptions about the presence of domain walls.

While there is evidence of edge states from planar tunneling in \sro,\cite{edgetunnel}, the fields due to edge currents have not been observed. This is perhaps the most striking signature of a chiral p-wave superconductor as it ties directly to the Majorana edge modes, the angular momenta of the Cooper pairs, and the macroscopic angular momentum, $N\hbar /2$, of an ideal chiral p-wave superconducting disc, in the absence of screening. The $\mu$SR experiments are interpreted as evidence for these currents at internal domain walls, so one would expect to also see them by scanning a SQUID loop over the $ab$\ surface, detecting the fields from any domain walls crossing, or coming within a penetration depth, of the  surface, as well as at the edges of the $ab$\ surface. Experiments now place an upper bound on these fields that is about 3 orders of magnitude below that expected from the simplest chiral p-wave models.\cite{kirtley,hicks2,budakian,curran} As mentioned previously, chiral p-wave models with large gap anisotropy on all 3 bands can reduce the currents by an order of magnitude, and even by 3 orders of magnitude if the edges are sufficiently disordered to be metallic.\cite{lederer,nontopo,largechern} Another possibility is facetted edges,\cite{retro,facet} although to give a current which is almost zero would require specific faceting on a length scale of $\sim$0.1 microns. Even with these possibilities, it is difficult to reconcile the positive $\mu$SR signal and the null surface current results with each other, suggesting further experiments and theoretical work to address this issue is required.

As discussed in subsection \ref{why}, chiral superconductivity occurs when the dominant order parameter is two-fold degenerate. By compressing the system along $\hat x$ in a $p_x \pm i p_y$ superconductor, one would expect to split the transition into two, the first at a slightly higher \tc\ into a $p_x$ superconducting state (assuming that compressing the system along $\hat x$ enhances $p_x$ superconductivity) and a lower transition where a second component $\pm i p_y$ grows up,  eliminating the nodes along $\hat y$.  The temperature of the upper transition should vary linearly with strain, and, as the strain passes through zero, there should be a cusp where the upper transition switches over to the $p_y$ phase.\cite{domain2,cusptheory}

Exactly this experiment was performed by Hicks and co-workers\cite{hicks3} who measured \tc\ as a function of unixial strain along the $\langle 100\rangle$\ and $\langle 110\rangle$\ directions for a \sro\ crystal.  What they saw was very different from expectations.  There was no sign of a cusp.  For the  $\langle 100\rangle$ direction the change in \tc\ was quadratic, positive and surprisingly large, increasing from about 1.4K to almost 2K for strains up to $\pm 0.2\%$.  In comparison, \tc\ in the $\langle 110\rangle$\ direction barely changed.  One possible interpretation is that the cusp exists but is very small and is overwhelmed by the surprisingly large quadratic dependence of \tc\ on strain, which may be related to the proximity of the Fermi energy to a van Hove singularity of the $\gamma$\ band.

Finally, the issue of low-lying excitations in the superconducting state of \sro\ should be mentioned. Numerous experiments show evidence of low-lying excitations  and the issue of where these are in momentum space has received considerable attention. Ref.~\onlinecite{Mackenzie} reviews this issue in some depth, pointing out most experiments are compatible with either vertical line nodes (along $k_z$ as one has in the d-wave curprates, for example) or horizontal line nodes but that experiments meant specifically to look for vertical line nodes had failed to confirm their existence. This issue is still unresolved. Recent STM experiments that tunnel into the $ab$\ surface, i.e. tunnelling along the $c$-axis, address this issue.\cite{davis}  Only one gap is observed and, since the $\gamma$\ band has almost no dispersion along the $c$-axis, it is reasonable to assume this is the gap on the $\alpha$\ and $\beta$\ bands. These experiments are consistent with either vertical line nodes or near-nodes on the $\alpha$\ and $\beta$\ bands and, consequently, are consistent with the weak-coupling RG results.\cite{scaffidi} They do not address the anisotropy of the $\gamma$\ band, which impacts most strongly on the magnitude of the chiral edge currents. This is another area where further work is needed.

\section{\upt}
\label{secUPt}

\upt\ is a heavy fermion metal which undergoes a double transition  from a metal to a non-chiral superconducting state at \tc$^+\approx$\ 0.53K and to a second superconducting state, thought to be chiral, at \tc$^-\approx\ $0.48K.\cite{jt}  The structure of \upt\ is hexagonal-close-packed with hexagonal layers of U and Pt ions forming \upt\ layers which are stacked in an ABAB sequence. Because nearest neighbor pairs of U ions in a layer are each separated by a Pt ion, the U ions are much farther apart than in U metal and conduction between the partially occupied U 5f orbitals proceeds via admixtures of unoccupied Pt orbitals.  Since the layers are closely spaced, transport in the $c$-direction is somewhat larger than in the planes, making this a fully 3D system.  

\begin{figure}
\includegraphics[width=0.8\textwidth]{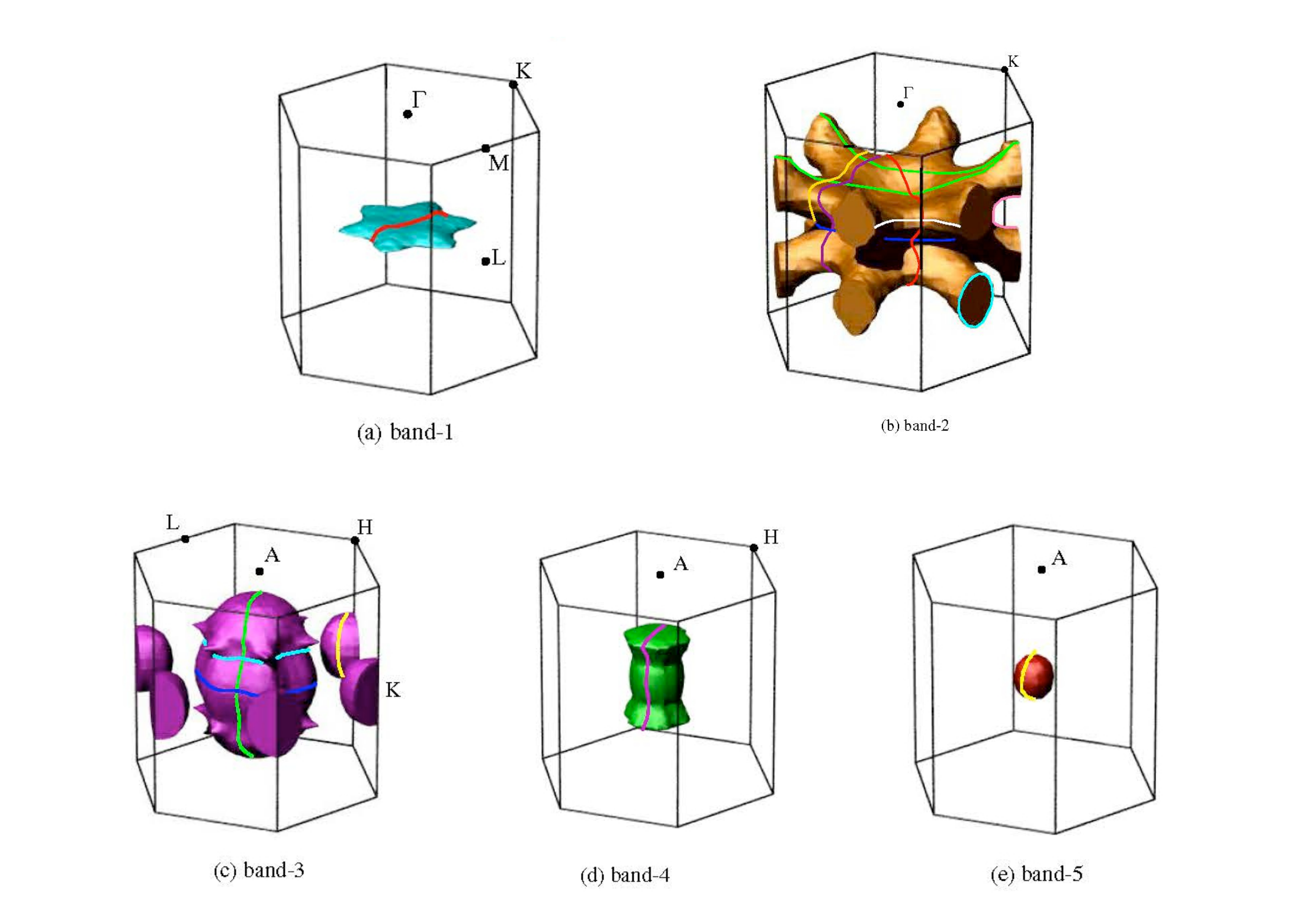}
\caption{\label{fsupt3} Fermi surfaces of \upt\ obtained from de Haas van Alphen measurements from Ref.~\onlinecite{julian}.  Surfaces (a) and (b) are hole Fermi surfaces centered on the A point, while surfaces, (c-e) are electron surfaces centered on the $\Gamma$ point. }
\end{figure}
Although its quasiparticles are heavy and the low temperature specific heat very large, the resistivity of \upt\ decreases monotonically with decreasing temperature, unlike most heavy fermion metals, and so at low temperatures it can be thought of as a Fermi liquid with strong electron-electron interactions.\cite{jt}
The fact that large extremely perfect crystals of \upt\ can be grown has made it possible to determine by de Haas van Alphen measurements\cite{dhva1, dhva2, julian} details of its complex Fermi surface that consists of several complicated sheets,  as can be seen in  Fig.~\ref{fsupt3}.

Band structure calculations show that these multiple sheets arise from a group of 5 or 6 bands crossing the Fermi energy.\cite{abc}  The bands are composed of U 5f orbitals hybridized with Pt orbitals.  Strong spin-orbit coupling splits these bands into j=5/2 and 7/2, with the latter raised to higher energy, so that the bands crossing the Fermi energy are predominantly Kramer's doublets of j=5/2 states, of which there are six because of the two U ions per unit cell.  In the superconducting state, all of these Fermi surfaces must be gapped.

The symmetry of the superconducting gap has been the subject of extensive theoretical discussion.  Experiments now point to a so-called E$_{2u}$ symmetry gap function involving nearly degenerate $(k_x^2-k_y^2)k_z$ and $k_xk_yk_z$ symmetry order parameters. Because of the odd spatial symmetry of the gap function, this superconducting state should have triplet pairing.  However the nature of this triplet state involves pairing of spins in states which are Kramers doublets, due to the strong spin-orbit coupling. Direct evidence for triplet pairing is rather slim.  It is true that the Knight shift changes rather little in the superconducting state,\cite{jt,tou1,tou2} but, given the strong spin-orbit coupling and multiple bands the interpretation is far from clear.

The two symmetry-related spatial order parameters of the superconducting state are believed to be weakly split by some perturbation such as weak antiferromagnetism or a small lattice distortion that favors one of the two and causes it to condense at the higher \tc.  At a slightly lower temperature, the second order parameter is thought to condense with a relative phase of $\pm i$, breaking time-reversal symmetry and chirality. The first transition results in a real order parameter with nodes in orthogonal directions in the $k_x$-$k_y$ plane.  At the second (lower) transition the system becomes fully gapped as the other order parameter grows up with relative phase $\pm i$.  The presence of a nodal surface at $k_z = 0$ leads to low-lying excitations that are observed in low-temperature transport.\cite{jt} 

Evidence for broken time-reversal symmetry was first seen in $\mu$SR experiments of Luke et al.\cite{luke}, who observed an onset of internal fields which broadened the zero field $\mu$SR resonance at the lower transition. However their results were disputed after measurements by Dalmas de R\'eotier et al.\cite{ddr} who saw no additional broadening at \tc\ in what were stated to be much higher quality samples. In principle, internal magnetic fields should only appear in a chiral superconductor due to spontaneous currents that are generated at chiral domain boundries and crystal defects and surfaces, as has been discussed above for the case of \sro.  One interpretation of the two contradictory $\mu$SR results is that chiral superconductivity in the higher quality samples of Dalmas de R\'eotier et al. grew up in a single domain and hence internal fields were not visible in their more perfect samples.  Support for this interpretation has been found recently in polar Kerr effect measurements, by the Stanford group of Kapitulnik,\cite{schemm} which showed a substantial rotation of the Kerr angle growing up below the lower transition. Unlike $\mu$SR, the Kerr effect is largest for single-domain chiral superconductivity and hence is most easily observed in the cleanest samples, assuming an intrinsic mechanism is responsible.

\begin{figure}
\includegraphics[width=0.8\textwidth]{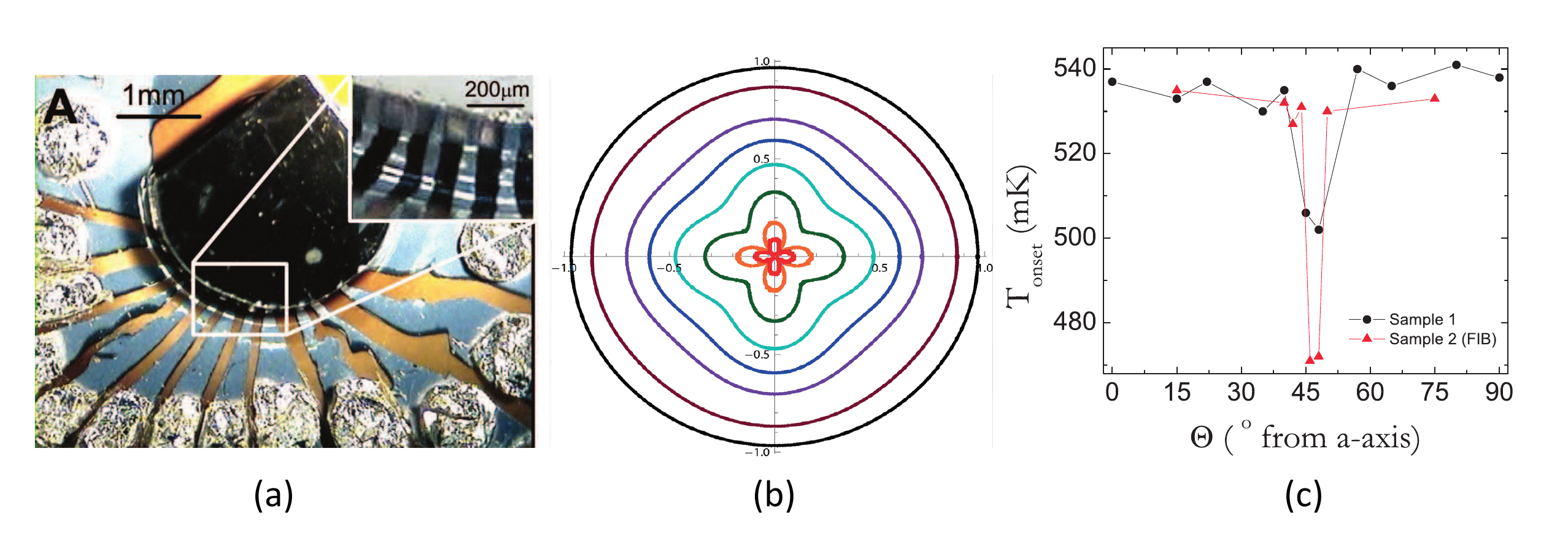}
\caption{\label{vhar} (a) Experimental configuration used in Ref.~\onlinecite{vhar} to measure the angular dependence of the critical current for a \upt\ crystal.  Multiple junctions are arranged uniformly around the edge of a \upt\ crystal covering an angle of about  90$^{\circ}$. 
(b) Schematic illustration of the temperature of the superconducting gap in \upt . The two innermost curves represent the gap for \tc $^-$ $<\ $ T $<\ $ \tc $^+$, 
while successively larger contours represent the development with decreasing temperature below \tc $^-$. 
(c) Onset temperature for non-zero critical current as a function of angle in the crystal.}
\end{figure}
Evidence for nodes in the angular dependence of the gap in the $k_x$-$k_y$ plane between the two transitions and for chiral symmetry breaking at the lower transition has been provided by angular-resolved Josephson tunneling experiments by the group of van Harlingen.\cite{vhar} 
In these measurements, an array of Josephson junctions is attached around the edge of a cylindrical wafer of \upt , as shown in Fig.~\ref{vhar}a, covering a range of about 90$^{\circ}$, where 0$^{\circ}$ is along the $\hat a$-axis direction in the hexagonal basal plane. Measuring the critical current as a function of angle and temperature, the authors found that the critical current turned on at the upper transition temperature, Fig.~\ref{vhar}c, except around 45$^{\circ}$ where the onset dipped down to the lower transition.  This is consistent with the temperature dependence of the gap function shown in Fig.~\ref{vhar}b, where a $k_x^2 - k_y^2$  gap (the innermost red curves) grows up at the higher transition, developing into a $k_x^2 - k_y^2\pm ik_xk_y$ gap below the lower transition. 

One might ask whether these results suggest that the gap is d-wave, instead of the expected triplet chiral f-wave with spatial dependence   $(k_x^2 - k_y^2\pm ik_xk_y)k_z$.  In the latter case, one might expect the tunneling current through an $a$-$c$ face to vanish because of the $k_z$ symmetry. The authors note however that the as-grown faces of the crystal have a surface normal tipped about 3$^{\circ}$ out of the basal plane, which might be enough to account for the non-zero critical current.  In any case, it is remarkabe that the critical current would exhibit four-fold symmetry in these nearly perfect hexagonal crystals.

Since experiments suggest that, unlike \sro, \upt\ crystals form a single chiral domain, it would be interesting to look for surface fields due to edge currents. These currents are expected to be suppressed compared to simple chiral p-wave because of the higher angular momentum.\cite{higherchirality} The small Fermi velocities in \upt\ further suppress the current. However, the above experiments are consistent with an isotropic gap in the $ab$\ or $k_a$-$k_b$\ plane and the samples are ultra-pure, two positive conditions, in addition to being a single domain, for observable edge currents.

\section{Other Possible Chiral Superconductors}
\label{sec7}

While \sro\ and \upt\  have been the most intensely studied candidates for chiral superconductivity, there are other materials which are possible candidates for chiral superconductivity. SrPtAs is an interesting material that is referred to as ``locally non-centrosymmetric'' because, although the crystal as a whole has inversion symmetry, the unit cell contains two inequivalent PtAs honeycomb layers, each of which lacks inversion symmetry. $\mu$SR measurements see TRSB below \tc$\sim$2.4K.\cite{biswas} With hexagonal structure, \mbox{SrPtAs} is a candidate for chiral d-wave order, although other states with BTRS are also allowed by symmetry in this unusual crystal structure.\cite{fischer}  While some data are consistent with either triplet or singlet pairing, \cite{bruckner} recent NMR measurements of the Knight shift and the penetration depth point toward singlet pairing with an isotropic (or nodeless) gap.\cite{matano} Functional RG studies on a microscopic model appropriate for SrPtAs find chiral d-wave superconductivity is favoured.\cite{rg-spa} Since the data mostly points toward singlet, nodeless superconductivity with BTRS, chiral d-wave order is a strong candidate. To date, experiments have been done on polycrystalline samples. If single crystals could be made this would open the door to more detailed studies of the superconducting order.

The heavy fermion, URu$_2$Si$_2$, is another fascinating material that has been intensely studied because it displays a mysterious ``hidden order'' phase. That is, it undergoes a transition at 17.5K, but the nature of the order in the low-temperature phase has long eluded identification, with recent work proposing, for example, nematic order\cite{okazaki}, density wave,\cite{timusk} chiral order,\cite{kung} and an unusual double time-reversal symmetry breaking.\cite{coleman} Below 1.4K, URu$_2$Si$_2$\ is superconducting.  Magnetic torque measurements see evidence for BTRS in the superconducting state,\cite{zhang} and a non-zero Kerr effect is observed.\cite{schemm2} The tetragonal crystal symmetry and the observation of BTRS, along with other measurements, point toward a chiral d-wave state.\cite{yano,kasahara,hsu}  Given the purity of the samples, it is thought the Kerr effect is likely intrinsic, perhaps resulting from the inter-band pairing mechanism.\cite{schemm2}  It has been suggested that the pairing mechanism may be quite novel.\cite{schemm2,coleman} The unusual nature of the hidden order state from which superconductivity condenses, together with a variety of other anomalous properties of this heavy fermion material, makes understanding the microscopic mechanism of superconductivity a particularly challenging problem and many open questions remain.

Another interesting system is the water doped colbaltates, Na$_x$CoO$_2\cdot y$H$_2$O, which have a superconducting dome near $x=0.3, y=1.3$ with a maximum \tc\ of 4.5K.\cite{co-dome} This is a layered material and the cobalt ions sit on a triangular lattice with oxygen in-between. This material is thought of as triangular lattice version of the high \tc\ cuprates. The hexagonal crystal structure and the fact that NMR points to singlet pairing, makes chiral d-wave superconductivity a possibility. In fact, a combined dynamical mean field and RG approach finds an anisotropic chiral d-wave state is stabilized over a range of doping, by a combination of multi-orbital effects, Fermi surface topology and magnetic fluctuations.\cite{thomale2} The chiral d-wave order found is highly anisotropic with near-nodes, which is consistent with experiments.\cite{thomale2} However, perhaps because of materials issues, many experiments have yet to be done on this material. In particular, we are not aware of experiments which find direct evidence of BTRS. Given the theoretical work, $\mu$SR and polar Kerr experiments to look for possible BTRS would be of great interest.

One of the most interesting proposals for chiral superconductivity is in doped graphene. This is a theoretical proposal since superconductivity has not been observed in graphene. However, an RG analysis predicted chiral d-wave superconductivity might be stabilized by repulsive interactions when single layer graphene is doped.\cite{graphene}  At either 3/8 or 5/8 filling (undoped graphene is at 1/2 filling), the Fermi surface has perfect nesting (neglecting further neighbor hopping) and the density of states is logarithmically divergent due to the Fermi surface coinciding with the van Hove points at these fillings.  Consequently the effect of interactions is strongly enhanced and, if superconductivity is the strongest instability, they find chiral d-wave is favoured.\cite{graphene}  Other theoretical work also suggests looking for chiral superconductivity in systems with a honeycomb lattice structure close to a Mott insulating phase.\cite{dwavehoneycomb,pwavehoneycomb}

Finally, we mention that the examples discussed here of materials which have been proposed as possible chiral superconductors do not exhaust the possibilities. In addition to other superconducting compounds that have shown some evidence of TRSB, one might also be able to stabilize chiral superconductivity or superfluidity in cold atom systems or by doping a topological insulator.\cite{qi-zhang} Furthermore, hybrid systems of topological insulators combined with an s-wave superconductor can behave analogously to a chiral superconductor  through the proximity effect and this is currently an active field of study.\cite{hybrid}  

\section{Final Thoughts}
\label{sec8}

Volovik has described the wealth of phenomena associated with chiral p-wave triplet superconductivity in a neutral superfluid as ``The Universe  in a Helium Droplet.''\cite{volovik}  Much of this rich complexity is the result of both chiral symmetry breaking and also the extra degrees of freedom associated with a spin triplet order parameter, which lead to phenomena such as spontaneous surface currents, Majorana fermions, half-quantum and other exotic vortices, and a variety of collective modes and textures.  The chiral superconductors discussed in this paper have the important additional elements of electric charge and charged currents, along with possibly interesting multi-orbital and spin-orbit effects which we are only beginning to understand. 

Modern nanofabrication techniques allow us to probe and manipulate chiral superconductors and to incorporate them into devices. For example, it has been proposed\cite{wire} that Majoranas might be induced at the ends of a quantum wire proximity-coupled to the surface of a chiral p-wave superconductor.  This would be a very interesting experiment to try with \sro. It would be analogous to the recent observations by Yazdani and co-workers\cite{yazdani} of Majorana end states for a chain of Fe atoms on a surface of (non-chiral) superconducting Pb. Such new phenomena, once they are well-understood and characterized, can be expected to lead to new applications.  

Unfortunately, we are not quite there yet because there is still uncertainty about important details.  The elusiveness of what should be ubiquitous surface currents is puzzling, although detailed theories are emerging which are consistent with these currents being small.  Much work remains to establish the existence of half-flux quantum vortices.  The magnetic susceptibility of real triplet superconductors needs to be better understood.  The existence of spontaneously broken time reversal symmetry seems clear from polar Kerr effect measurements, but it is not yet clear whether or how other probes of spontaneous symmetry breaking are obscured by chiral domains. One would also like to have detailed information of where the low-lying excitations in \sro\ are in momentum space and, in all the candidate materials, detailed information on the  gap structure on different bands.  There is plenty of room for clever new experiments to probe the distinctive properties of chiral superconductors in novel ways, as well as for the discovery of new materials which exhibit these properties.
\begin{acknowledgements}
We are indebted to Clifford Hicks, Wen Huang, John Kirtley, Samuel Lederer, Kathryn Moler, Srinivas Raghu,  Edward Taylor and Gertrud Zwicknagl for collaborations on work related to chiral superconductivity. We also acknowledge support from NSERC, CIFAR, the Canada Council Killam Program and the NSF under Grant No. NSF PHY11-25915.

\end{acknowledgements}
%__________________________________________________________


\begin{thebibliography}{99}
%-Sec 1
\bibitem{ko} Kamerlingh Onnes H 1911 (Nov) {\it Commun. Phys. Lab. Univ. Leiden. Suppl.}~\textbf{29} 
\bibitem{bcs} Bardeen J, Cooper L N and Schrieffer J R 1957 {\it Phys.~Rev.} ~\textbf{108} 1175
\bibitem{parks} For an overview of the status of superconductivity theory and experiment in 1969, see \textit{Superconductivity (in Two Volumes)}, ed. R.~D.~Parks, Marcel Dekker, Inc.~New York, N.~Y.~ 1969
\bibitem{am} Anderson P W and Morel P 1961 {\it Phys. Rev.} \textbf{123} 1911
\bibitem{bw} Balian R and Werthamer N R {\it Phys. Rev.} \textbf{131} 1553
\bibitem{orl} Osheroff D D, Richardson R C and Lee DM 1972 {\it Phys.~Rev.~Lett.} \textbf{28}8 85
\bibitem{leggett} For an oveview of the theory of superfluid $^3$He, see Leggett A J 1975 {\it Rev. Mod. Phys.} \textbf{47} 331
\bibitem{leggettbook} Legget A J 2006 \textit{Quantum Liquids: Bose Condensation and Cooper Pairing in Condensed-Matter Systems (Oxford Graduate Texts)}
\bibitem{wheat} For an overview of early experiments on superfluid $^3$He, see Wheatley J C 1975 {\it Rev. Mod. Phys.} \textbf{47} 415
\bibitem{bm} Bednorz J G and Muller K A 1986 {\it Z.~Physik} \textbf{B 64} 189
\bibitem{beenakker} Beenakker C W J 2013 {\it Ann.~Rev.~Cond.~Mat.~Phys.}  \textbf{4}113
\bibitem{ivanov} Ivanov D A 2001 {\it Phys. Rev. Lett}.\textbf{ 86} 268 
\bibitem{npj} Das Sarma S, Freedman M and Nayak C 2015 {\it npj|Quantum Information} \textbf{1} 15001; arXiv:1501.02813 
\bibitem{chiraltop} See Schnyder A P, Ryu S, Furusaki A and Ludwig A W W2008 {\it Phys. Rev.} B \textbf{78} 195125 and references therein.
\bibitem{read} Read N and Green D 2000 {\it Phys. Rev.} B \textbf{61} 10267
\bibitem{stone} Stone M and Roy R 2004 {\it Phys. Rev.} B \textbf{69} 184511
\bibitem{senthil} Senthil T, Marston J B and Fisher M P A 1999 {\it Phys. Rev.} B \textbf{60} 4245
\bibitem{newsigrist} Imai Y, Wakabayashi K and Sigrist M 2016 \textit{Phys. Rev.} B \textbf{93} 024510 
\bibitem{maenotriplet} Maeno Y, Kittaka S, Nomura T, Yonezawa S, Ishida K 2012 {\it  J~ Phys.~Soc.~Jpn.} \textbf{81} 011009
\bibitem{budakian} Jang J, Ferguson D G, Vakaryuk V, Budakian R, Chung S B, Goldbart P M, Maeno Y 2011 {\it Science} \textbf{331} 186
\bibitem{dwave} Kirtley J R, Tsuei C C, Sun J Z, Chi C C, Yu-Jahnes L S, Gupta A, Rupp M, and Ketchen M B 1995 \textit{Nature} \textbf{373} 225-228
\bibitem{domain1} Volovik G E and  Gorkov L P  1985 \textit{Sov. Phys.-JETP} \textbf{61} 843
\bibitem{domain2}  Sigrist M and Ueda K 1991 \textit{Rev. Mod. Phys.} \textbf{63} 239
%-Sec 2
\bibitem{soc1} Haverkort M W, Elfimov I S, Tjeng L H, Sawatzky G A and Damascelli A 2008 {\it  Phys. Rev. Lett.} \textbf{101} 026406 
\bibitem{soc2} Rozbicki E J, Annett J F, Souquet J R and Mackenzie A P 2011 {\it J. Phys. Cond. Matt.} \textbf{23} 094201
\bibitem{abc} Albers R C, Boring A M and Christensen N E 1986 \textit{Phys. Rev.} B \textbf{33} 8116
%-Sec 3
\bibitem{kitaev} Kitaev A Y \textit{Chernogolovka 2000: Mesoscopic and strongly correlated electron systems, Usp. Fiz. Nauk (Suppl.)} \textbf{171} 132
\bibitem{gurarie} Gurarie V and Radzihovsky L 2007 {\it Phys. Rev.} B\textbf{75} 212509
\bibitem{niu} Niu Y, Chung S B, Hsu C H, Mandal I, Raghu S, and Chakravaty S {\it Phys. Rev.} B \textbf{85} 035110
\bibitem{nontopo} Huang W, Lederer S, Taylor E and Kallin C 2015 {\it Phys. Rev.} B \textbf{91} 094507
\bibitem{higherchirality} Huang W, Taylor E and Kallin C 2014 {\it Phys. Rev.} B \textbf{90} 224519
\bibitem{kita} Kita T 1998 {\it J. Phys. Soc. Jap.} \textbf{67} 216
\bibitem{volovik1} Volovik G E 1988 {\it Phys. Scr.} \textbf{38} 321; 1988 {\it Zh. Eksp. Teor. Fiz.} \textbf{94} 123 (English translation: 1988 {\it Soviet Physics (JETP)} \textbf{67} 1804)

\bibitem{oshikawa} Tada Y, Nie W and Oshikawa M 2015 {\it Phys. Rev. Lett.} \textbf{114} 195301
\bibitem{largechern}  Scaffidi T and Simon S H 2015 {\it Phys. Rev. Lett.} \textbf{115} 087003
\bibitem{retro}  Sauls J A 2011 {\it Phys.Rev. B} \textbf{84} 214509
\bibitem{facet} Bouhon A and Sigrist M 2014 {\it Phys. Rev.} B \textbf{90} 220511(R) 
\bibitem{lederer} Lederer S, Huang W, Taylor E, Raghu S and Kallin C 2014 {\it Phys. Rev.} B \textbf{90} 134521
\bibitem{matsumoto} Matsumoto M and Sigrist M 1999 {\it J. Phys. Soc. Jpn.} \textbf{68} 994 


\bibitem{tewari} Tewari S, Das Sarma S, Nayak C, Zhang C and Zoller P 2007 {\it Phys.~Rev.~Lett.} \textbf{98} 010506
%\bibitem{kopnin} Kopnin N B and Salomaa M M 1991 {\it Phys. Rev.} B \textbf{44} 9667
\bibitem{bauer} Bauer B, Lutchyn R M, Hastings M B and Troyer M 2013 {\it Phys. Rev.} B \textbf{87} 014503
\bibitem{sukbum} Chung S B, Bluhm H and Kim E A 2007 {\it Phys. Rev. Lett.} \textbf{99} 197002


%-Sec 4 have checked this and later sections for no et als for 10 or less

\bibitem{roy} Roy R and Kallin C 2008 {\it Phys. Rev.} B \textbf{77} 174513
\bibitem{cupratekerr} Hosur P, Kapitulnik A, Kivelson S A, Orenstein J, Raghu S, Cho W and Fried A 2015 \textit{Phys. Rev.} B \textbf{91} 039908
\bibitem{goryo99} Goryo J and Ishikawa K 1999 {\it Phys. Lett.} A \textbf{260} 294
\bibitem{golub} Horovitz B and Golub B 2002 {\it Europhys. Lett.} \textbf{57} 892
\bibitem{goryo} Goryo J 2008 {\it Phys. Rev.} B \textbf{78} 060501(R)
\bibitem{yakovenko} Lutchyn R M, Nagornykh P and Yakovenko V M 2009 {\it Phys. Rev.} B \textbf{80} 104508
\bibitem{spivak} Li S, Andreev A V and Spivak BZ 2015 {\it Phys. Rev.} B \textbf{92} 100506
\bibitem{yip} Yip S K and Sauls J A 1992 {\it J. Low Temp. Phys.} \textbf{86} 257
\bibitem{taylor} Taylor E and Kallin C 2012 {\it Phys. Rev. Lett.} \textbf{108} 157001
\bibitem{taylor2} Taylor E and Kallin C 2013 {\it J. Phys. Conf. Ser.} \textbf{449} 012036
\bibitem{annett1} Wysokinski K, Annett J F and Gyorffy B L 2012 {\it Phys. Rev. Lett.} \textbf{108} 077004
\bibitem{annett2} Gradhand M, Wysokinski K K, J.F. Annett and Gyorffy B L 2013 {\it Phys. Rev.} B \textbf{88} 094504

%-Sec 5
\bibitem{maeno94} Maeno Y, Hashimoto H, Yoshida K, Nishizaki S, Fujita T, Bednorz J and Lichtenberg F 1994 {\it Nature} \textbf{372} 532
\bibitem{rice95} Rice T M and Sigrist M 1995 {\it J. Phys. Cond. Matt.} \textbf{7} L643
\bibitem{baskaran} Baskaran G 1996 {\it Physica} B \textbf{223-224} 490
\bibitem{puzzles} Kallin C and Berlinsky A J 2009 {\it J. Phys. Cond. Matt.} \textbf{21} 164210
\bibitem{Mackenzie} Mackenzie A P and Maeno Y 2003 {\it Rev. Mod. Phys.} \textbf{75} 657 
\bibitem{kallin} Kallin C 2012 {\it Rep. Prog. Phys.} \textbf{75} 042501

\bibitem{damascelli} Damascelli A et al. 2000 {\it Phys Rev. Lett.} \textbf{85} 5194 
\bibitem{soc} Yanase and Ogata M 2003 {\it J. Phys. Soc. Jpn.} \textbf{72} 673

\bibitem{spinarpes} Veenstra C N, Zhu Z H, Raichle M, Ludbrook B M, Nicolaou A, Slomski B, Landolt G, Kittaka S, Maeno Y, Dil J H, Elfimov I S, Haverkort M B, and  Damascelli A, 2014 {\it Phys. Rev. Lett.} {\bf 112} 127002
\bibitem{disorder} Mackenzie A P, Haselwimmer R K W, Tyler A W, Lonzarich G G, Mori Y, Nishizaki S and Maeno Y 1998 {\it Phys. Rev. Lett.} \textbf{80} 161; Erratum {\it ibid.} 3890 
\bibitem{phonon} Mao Z Q, Maeno Y, Mori Y, Sakita S, Nimori S and Udagawa M 2001 {\it Phys. Rev.} B \textbf{63} 144514
\bibitem{dagotto} Dagotto E, Hotta T and Moreo A 2001 {\it Phys. Rep.} \textbf{344} 1 
\bibitem{raghu} Raghu S, Kapitulnik A and Kivelson S A 2010 {\it Phys. Rev. Lett.} \textbf{105} 136401 
\bibitem{frg} Wang Q H, Platt C, Yang Y, Honerkamp C,  Zhang F C, Hanke W, Rice T M and Thomale R 2013 {\it Euro. Phys. Lett.} \textbf{104} 17013
\bibitem{scaffidi} Scaffidi T, Romers J C and Simon S H 2014 {\it Phys. Rev.} B \textbf{89} 220510
\bibitem{davis} Firmo I A, Lederer S, Lupien C, Mackenzie A P, Davis J C and Kivelson S A 2013 {\it Phys. Rev.} B \textbf{88} 134521
\bibitem{ohno} Tsuchiizu M, Yamakawa Y, Onari S, Ohno Y and Kontani H 2015 {\it Phys. Rev.} B \textbf{91} 155103
\bibitem{puetter} Puetter C M and Kee H Y 2010 {\it Euro. Phys. Lett.} \textbf{98} 27010
\bibitem{annett}  Annett J F, Litak G, Gyorffy B L and Wysokinski K I 2002 {\it Phys. Rev.} B \textbf{66} 134514

\bibitem{mazin} Zutic I and Mazin II 2005 {\it Phys. Rev. Lett.} \textbf{95} 217004 
\bibitem{nmr1} Ishida K,  Mukuda H,  Kitaoka Y, Asayama K, Mao Z Q,  Mori Y and Maeno Y 1998 {\it Nature} \textbf{396} 658
\bibitem{nmrc} Murakawa H, Ishida K, Kitagawa K, Mao Z Q and Maeno Y 2004 {\it Phys. Rev. Lett.} \textbf{93} 167004
\bibitem{neutron} Duffy J A, Hayden S M, Maeno Y, Mao Z, Kulda J and McIntyre G J 2000 {\it  Phys. Rev. Lett.} \textbf{85} 5412
\bibitem{liu} Nelson K D, Mao Z Q, Maeno Y and Liu Y 2004 {\it Science} {\it 306} 1151
\bibitem{imada-new} Ishida K, Manago M, Yamanaka T, Fukazawa H, Mao Z Q, Maeno Y and Miyake K 2015 {\it Phys. Rev.} B \textbf{92} 100502(R)
\bibitem{theory} Miyake K 2014 {\it J. Phys. Soc. Jap.} \textbf{83} 053701
\bibitem{drotation} Annett J F, Gyorffy B L, Litak G and Wysokinski K I 2008 {\it Phys. Rev.} B \textbf{78} 054511
\bibitem{Pavarini} Pavarini E and Mazin I I 2006 {\it Phys. Rev.} B \textbf{74} 035115 
\bibitem{roberts} Roberts K, Budakian R and Stone M 2013 {\it Phys. Rev.} B \textbf{88} 094503
\bibitem{xcai} Cai X, Ying Y A, Staley N E, Xin Y, Fobes D, T J Liu, Mao Z Q, Liu Y 2013 {\it Phys. Rev.} B \textbf{87} 081104(R)
\bibitem{musr1} Luke G M et al. 1998 {\it Nature} \textbf{394} 558
\bibitem{musr2} Luke G M and Sonier J 2011 {\it Phys. in Canada} \textbf{67} 93
\bibitem{kapitulnik} Xia J, Maeno Y, Beyersdorf P T, Fejer M M and Kapitulnik A 2006 {\it Phys. Rev. Lett.} \textbf{97} 167002
\bibitem{wollman} Wollman D A, Van Harlingen D J, Lee W C, Ginsberg D M and Leggett A J 1993 {\it Phys. Rev. Lett.} \textbf{71} 2134
\bibitem{dale} Kidwingira F, Strand J D, van Harlingen D J and Maeno Y 2006 {\it Science} \textbf{314} 1267
\bibitem{sigrist-domain} Bouhon A and Sigrist M 2010 {\it New J.~Phys.} \textbf{12} 043031
\bibitem{edgetunnel} Kashiwaya S, Kashiwaya H, Kambara H, Furuta T, Yagushi H, Tanaka Y and Maeno Y 2011 {\it Phys. Rev. Lett.} \textbf{107} 077003 
\bibitem{kirtley} Kirtley J R, Kallin C, Hicks C, Kim E A, Moler K A and Maeno Y 2007 {\it Phys. Rev.} B \textbf{76} 014526
\bibitem{hicks2} Hicks C W, Kirtley J R, Lippman T M, Koshnick N C, Huber M E, Maeno Y, Maple M B and Moler K A 2010 {\it Phys. Rev.} B \textbf{81} 214501
\bibitem{curran} Curran P J, Bending S J, Desoky W M, Gibbs A S, Lee S L and Mackenzie A P 2014 {\it Phys. Rev.} B \textbf{89} 144504
\bibitem{cusptheory} Walker M B and Contreras P 2002 {\it Phys. Rev.} B \textbf{66} 214508 
\bibitem{hicks3} Hicks C W, Brodsky D O, Yelland E A, Gibbs A S, Bruin J A N, Nishimura K, Yonezawa S, Maeno Y and Mackenzie A P 2014 {\it Science} \textbf{344} 283 

%-Sec 6 - UPt3
\bibitem{jt} For a detailed review of the properties of \upt\, see Joynt R and Taillefer L 2002 {\it Rev. Mod. Phys.}  \textbf{74} 235 
\bibitem{dhva1} Taillefer L, Newbury R, Lonzarich G G, Fisk Z and Smith J L 1986 {\it J.~Magn.~Magn.~Mater.} \textbf{63-64} 372 
\bibitem{dhva2} Taillefer L and Lonzarich G G 1988 {\it Phys.~Rev.~Lett.} \textbf{60} 1570 
\bibitem{julian} McMullan G J, Rourke P M C, Norman M R, Huxley A D, Doiron-Leyraud N, Flouquet J, Lonzarich G G, McCollam A and Julian S R 2008 {\it New J. Phys.}  \textbf{10} 053029 

\bibitem{tou1} Tou H, Kitaoka Y, Kimura N, Onuki Y, Yamamoto E and
Maezawa K 1996 \textit{Phys. Rev. Lett.} \textbf{77} 1374
\bibitem{tou2} Tou H, Kitaoka Y, Ishida K, Asayama K, Kimura N,
Onuki Y, Yamamoto E, Haga Y and Maezawa K 1998 \textit{Phys.
Rev. Lett.} \textbf{80} 3129
\bibitem{vhar}  Strand J D, Bahr D J, van Harlington D J, Davis J P , Gannon W J and  Halperin W P 2010 {\it Science} \textbf{328} 1368 (2010)
\bibitem{luke} Luke G M, Keren A, Le L P, Wu W D, Uemura Y J, Bonn D A, Taillefer L and Garrett J D 1993 {\it Phys.~Rev.~Lett.} \textbf{71} 1466 
\bibitem{ddr} de Reotier P D, Huxley A, Yaouanc A, Flouquet J, Bonville P, Imbert P,  Pari P, Gubbens P C M and Mulders A M 1995 {\it  Phys.~Lett.} A \textbf{205} 239 
\bibitem{schemm}  Schemm E R, Gannon W J, Wishne C M, Halperin W P and Kapitulnik A 2014 {\it Science} \textbf{345} 190

%-Sec 7 - Other possible chiral superconductors

\bibitem{biswas} Biswas P K 2013 {\it Phys. Rev.} B \textbf{87} 180503
\bibitem{fischer} Fischer M H and Goryo J 2015 {\it J. Phys. Soc. Jpn.}  \textbf{84} 054705  
\bibitem{matano} Matano K, Arima K, Maeda S, Nishikubo Y, Kudo K, Nohara M and Zheng G 2014 {\it Phys. Rev.} B \textbf{89} 140504(R) 
\bibitem{bruckner}  Bruckner F, Sarkar R, Gunther M, Kuhne H, Luetkens H, Neupert T,  Reyes A P, Kuhns P L, Biswas P K, Sturzer T, Johrendt D and  Klauss H H 2014 {\it  Phys. Rev.} B \textbf{90} 220503 
\bibitem{rg-spa}   Fischer M H, Neupert T, Platt C, Schnyder A P, Hanke W, Goryo J, Thomale R, Sigrist M 2014 {\it Phys. Rev.} B \textbf{89} 020509 
\bibitem{okazaki}  Okazaki R, Shibauchi T, Shi H J, Y Haga, Matsuda T D, Yamamoto E, Onuki Y, Ikeda H and Matsuda Y 2011 {\it Science} \textbf{331} 439  
\bibitem{timusk} Hall J S and Timusk T 2014 {\it Phil. Mag.} \textbf{94} 3760 
\bibitem{kung}  Kung H H, Baumbach R E, Bauer E D, Thorsmølle V K, Zhang W L, Haule K, Mydosh J A and Blumberg G 2015 {\it Science} \textbf{237} 1339 
\bibitem{coleman} Chandra P, Coleman P and Flint R 2015 {\it Phys. Rev.} B \textbf{91} 205103 
\bibitem{zhang} Li G et al. 2013 {\it  Phy. Rev.} B \textbf{88} 134517 
\bibitem{schemm2} Schemm E R, Baumbach R E, Tobash P H, Ronning F, Bauer E D and Kapitulnik A 2015 {\it Phys. Rev.} B \textbf{91} 140506 
\bibitem{yano} Yano K, et al. 2008 {\it Phys. Rev. Lett.} {\bf 100} 017004
\bibitem{kasahara} Kasahara Y,  Shishido H, Shibauchi T, Haga Y, Matsuda T D, Onuki Y and Matsuda Y 2009 {\it New J. Phys.} \textbf{11} 055061
\bibitem{hsu} Hsu C H and Chakravarty S 2014 {\it Phys. Rev.} B \textbf{90} 134507; Chakravarty S and Hsu C H 2015 {\it Mod. Phys. Lett.} B \textbf{29} 1540053
\bibitem{co-dome} Takada K, Sakurai H, Takayama-Muromachi E, Izumi F, Dilanian A and Sasaki T 2003 {\it Nature} \textbf{422} 53
\bibitem{thomale2} Kiesel M, Platt C, Hanke W and Thomale R 2013 {\it Phys. Rev. Lett.} \textbf{111} 097001
\bibitem{graphene}  Nandkishore R, Levitov L S and Chubukov A V 2014 {\it Nature Physics}  \textbf{8} 158
\bibitem{pwavehoneycomb} Gu Z C, Jiang H C and Baskaran G  arXiv:1408.6820
\bibitem{dwavehoneycomb} Gu Z C, Jiang H C, Sheng D N, Yao H, Balents L and Wen X G 2013 {\it Phys. Rev.} B \textbf{88} 155112
\bibitem{qi-zhang} Qi X L and Zhang S C 2011 {\it Rev. Mod. Phys.} \textbf{83} 1057
\bibitem{hybrid}  See, for example, Li Z Z, Zhang F C and Wang Q H 2014 {\it Nature Sci. Rep.} \textbf{4} 6363 and references therein.

%-Sec 8 - Conclusions

\bibitem{volovik} Volovik G E 2003 The Universe in a Helium Droplet, Clarendon Press, Intnl. Series of Monographs on Physics

\bibitem{yazdani} Nadj-Perge S,  Drozdov I K, Li J, Chen H, Jeon S, Seo J, MacDonald A H, Bernevig B A and Yazdani A 2014 {\it Science} \textbf{346} 602

\bibitem{wire} Nakosai S, Budich J C, Tanaka Y, Trauzettel B and Nagaosa N 2013 {\it Phys. Rev. Lett.} \textbf{110} 117002

\end{thebibliography}
\end{document}